\documentclass[acmlarge,nonacm]{acmart}
\makeatletter
\newcommand{\myconfshort}{\acmConference@shortname}
\newcommand{\myconffull}{\acmConference@name}
\newcommand{\myconfdate}{\acmConference@date}
\newcommand{\myconfloc}{\acmConference@venue}
\AtBeginDocument{
  \fancypagestyle{firstpagestyle}{
    \fancyhead{}%
    \fancyfoot[C]{}%
  }
  \fancyhf{}
  \fancyhead[LO]{\@headfootfont\shorttitle}%
  \fancyhead[RE]{\@headfootfont\@shortauthors}%
  \fancyhead[LE]{\@headfootfont\footnotesize \myconfshort, \myconfdate, \myconfloc}%
  \fancyhead[RO]{\@headfootfont\footnotesize \myconfshort, \myconfdate, \myconfloc}%
  \fancyfoot[C]{}%
}
\makeatother
\acmBooktitle{\conffull\@ (\confshort), \confdate, \confloc}

\authorsaddresses{}  %

\copyrightyear{2026}
\acmYear{2026}
\setcopyright{cc}
\setcctype{by}
\acmConference[FAccT '26]{The 2026 ACM Conference on Fairness, Accountability, and Transparency}{June 25--28, 2026}{Montreal, QC, Canada}
\acmBooktitle{The 2026 ACM Conference on Fairness, Accountability, and Transparency (FAccT '26), June 25--28, 2026, Montreal, QC, Canada}
\acmDOI{10.1145/3805689.3806728}
\acmISBN{979-8-4007-2596-8/2026/06}

\usepackage[utf8]{inputenc}
\usepackage{natbib}
\usepackage{graphicx}
\usepackage{amsmath}
\usepackage{amsfonts}
\usepackage{hyperref}
\usepackage{booktabs}
\usepackage{xcolor}
\usepackage{float}
\usepackage{cleveref}
\usepackage{colortbl}

\usepackage[skins, breakable]{tcolorbox} %
\usepackage{subcaption}
\usepackage{multicol} %
\usepackage{tikz} %
\usetikzlibrary{
  arrows.meta,
  positioning,
  shapes.misc,
  shapes.geometric,
  calc,
  scopes,
  decorations.pathreplacing,
  fit,
  backgrounds
}

\usepackage{listings}
\usepackage{soul}

\definecolor{chat_yellow}{HTML}{DDAA33}
\definecolor{browser_blue}{HTML}{004488}
\definecolor{enterprise_red}{HTML}{BB5566}

\newcommand{\ulchat}[1]{\setul{1pt}{1.5pt}\setulcolor{chat_yellow}\ul{#1}}
\newcommand{\ulbrowser}[1]{\setul{1pt}{1.5pt}\setulcolor{browser_blue}\ul{#1}}
\newcommand{\ulent}[1]{\setul{1pt}{1.5pt}\setulcolor{enterprise_red}\ul{#1}}

\newtcolorbox{outlinebox}[1][red]{
  breakable,
  enhanced,
  frame hidden,
  colback=white,
  boxsep=0pt,
  left=0pt,
  right=0pt,
  top=0pt,
  bottom=0pt,
  parbox=false,
  borderline west={1pt}{-5pt}{#1},
  borderline east={1pt}{-5pt}{#1}
}

\begin{document}

\title[The 2025 AI Agent Index]{{\fontsize{22}{36}\selectfont The 2025 AI Agent Index}\\{\fontsize{13}{36}\selectfont Documenting Technical and Safety Features of Deployed Agentic AI Systems}}

\author{Leon Staufer}
\email{lets2@cam.ac.uk}
\authornote{Corresponding author <lets2@cam.ac.uk>}
\affiliation{%
  \institution{University of Cambridge}
  \city{Cambridge}
  \country{United Kingdom}
}

\author{Kevin Feng}
\authornotemark[2]
\affiliation{%
  \institution{University of Washington}
  \city{Seattle}
  \state{Washington}
  \country{USA}
}
\author{Kevin Wei}
\authornote{Equal contribution, randomized order.}
\affiliation{%
  \institution{Harvard Law School}
  \city{Cambridge}
  \state{Massachusetts}
  \country{USA}
}
\author{Luke Bailey}
\authornotemark[2]
\affiliation{%
  \institution{Stanford University}
  \city{Stanford}
  \state{California}
  \country{USA}
}
\author{Yawen Duan}
\authornotemark[2]
\affiliation{%
  \institution{Concordia AI}
  \country{China}
}
\author{Mick Yang}
\authornotemark[2]
\affiliation{%
  \institution{University of Pennsylvania}
  \city{Philadelphia}
  \state{Pennsylvania}
  \country{USA}
}
\author{A. Pinar Ozisik}
\authornotemark[2]
\affiliation{%
  \institution{Massachusetts Institute of Technology}
  \city{Cambridge}
  \state{Massachusetts}
  \country{USA}
}

\author{Stephen Casper}
\authornote{Co-senior author.}
\affiliation{%
  \institution{Massachusetts Institute of Technology}
  \city{Cambridge}
  \state{Massachusetts}
  \country{USA}
}

\author{Noam Kolt}
\authornotemark[3]
\affiliation{%
  \institution{Hebrew University of Jerusalem}
  \city{Jerusalem}
  \country{Israel}
}

\renewcommand{\shortauthors}{Staufer et al.}

\begin{abstract}
Agentic AI systems are increasingly capable of performing professional and personal tasks with limited human involvement.
However, tracking these developments is difficult because the AI agent ecosystem is complex, rapidly evolving, and inconsistently documented, posing obstacles to both researchers and policymakers.
To address these challenges, this paper presents the 2025 AI Agent Index.
The Index documents information regarding the origins, design, capabilities, ecosystem, and safety features of 30 state-of-the-art AI agents based on publicly available information and email correspondence with developers. 
In addition to documenting information about individual agents, the Index illuminates broader trends in the development of agents, their capabilities, and the level of transparency of developers. 
Notably, we find that transparency varies substantially across agent developers and observe that most developers share little information about safety, evaluations, and societal impacts. 
The 2025 AI Agent Index is available online at \url{https://aiagentindex.mit.edu}.
  
\end{abstract}


\maketitle

\section{Introduction}

Despite growing interest and investment in agentic AI systems capable of automating complex tasks with limited human involvement \citep{mckinsey2025agents, Pandey2025Agents,Singla2025StateAI,ontoforce2025agentic, Gibney2025AIAgents, gottweis2025towards, gridach2025agentic, wei2025ai}, key aspects of their real-world development and deployment remain opaque, with little information made publicly available to researchers or policymakers \citep{casper2025ai}. 
In particular, there are currently no clear answers to several basic questions concerning agentic AI systems:
\begin{itemize}
    \item Who is developing the most impactful agentic systems?
    \item In which domains are they deployed?
    \item What processes and resources are used to develop these systems?
    \item How are they evaluated?
    \item What guardrails are in place to mitigate their unique risks?
\end{itemize}

\begin{figure}
    \centering
    \includegraphics[width=\linewidth]{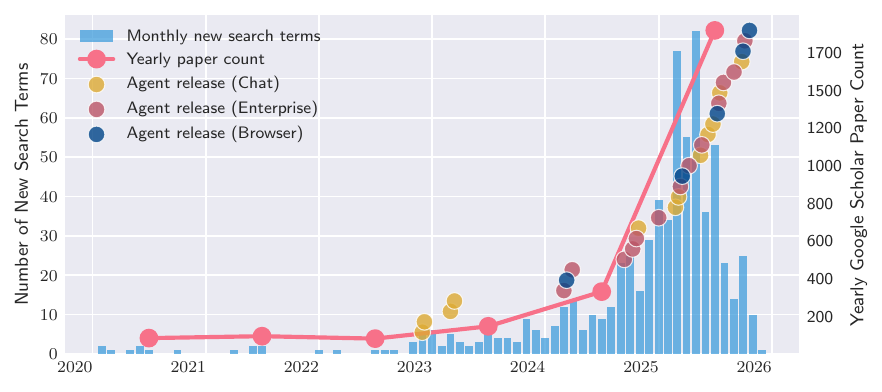}
    \Description{Composite chart of agent interest and releases over time. The horizontal axis runs from 2020 to early 2026; the left vertical axis shows monthly new search terms and the right vertical axis shows yearly Google Scholar paper counts. Blue bars rise from near zero before 2023 to a peak around mid to late 2025, while the red line of paper counts climbs sharply in 2025. Colored dots marking cumulative releases show chat agents appearing first, then enterprise and browser agents, with all categories accelerating in 2025.}
    \caption{\textbf{Interest in AI agents is growing.} 2025 has seen a sharp increase in interest in AI agents. This is reflected in an increase of new Google search terms related to agentic AI products (blue bars) as well as Google Scholar paper counts for ``AI agent'' or ``agentic AI'' (red line). Accumulation of individual releases of agentic AI products included in this Index is shown by category: \ulchat{chats with agentic tools}, \ulent{enterprise agents}, and \ulbrowser{browser agents}. See \Cref{fig:agent_release_timeline} for details on releases and \Cref{app:search} for details on public interest.}
    \label{fig:gs_searches}
\end{figure}

To answer these questions, we introduce and release the 2025 AI Agent Index. 
The Index provides in-depth information on 30 agentic systems across 6 categories: \ul{legal}, \ul{technical capabilities}, \ul{autonomy \& control}, \ul{ecosystem interaction}, \ul{evaluation}, and \ul{safety}. By focusing on the most widely deployed agents, the Index prioritizes depth over breadth: these systems are likely to have the greatest ecosystem impact, though emerging and smaller-scale agents may exhibit different patterns (see \Cref{sec:limitations}).
This 2025 Index follows the first 2024 AI Agent Index \citep{casper2025ai}.
To account for recent growth and change in the AI agent ecosystem (see \Cref{fig:gs_searches}), this 2025 Index develops and implements substantially revised inclusion criteria (\Cref{sec:inclusion_criteria}) and information fields (\Cref{sec:constructing}). 

In addition to providing information about prominent AI agents, this Index also reveals ecosystem-wide trends regarding which information developers do and do not publicly share. 
This sheds light on the state of transparency in the agent ecosystem amidst \href{https://incidentdatabase.ai/apps/discover/}{agentic AI incidents}, recent attention from governments \citep{oueslati2025ahead, usaisitechnicalstaff2025, bazinska2025breaking, korbak2025evaluate, zou2025security}, industry self-regulation efforts \citep{metr_faisc}, and gaps between expectations of agent developers and reality \citep{belcic2025aiagents}.
Among the key findings: most safety-related fields (135/240) have no public information available; nearly all indexed agents rely on just three foundation model families (GPT, Claude, Gemini); most agents do not disclose their AI nature to end users or third parties by default; and only four agents provide agent-specific safety evaluations.
We make three contributions:
\begin{enumerate}
    \item \textbf{Agent Index:} We index 30 highly agentic and widely used products (\Cref{sec:constructing}).\footnote{We use terms like ``agentic,'' ``pursue,'' and ``choose'' as shorthand for computational processes without attributing human-like intentionality, consciousness, or agency to AI systems. We recognize that such terms may anthropomorphize AI systems in a misleading way and obscure the sociotechnical nature of these systems \citep{barrowAnthropomorphismAIHype2024, inieAIProbabilisticAutomation2024}. When speaking of ``autonomy'' we only refer to technical automation without human-in-the-loop rather than independent volition. See \Cref{sec:background} for further discussion of the term ``agent''.}
    \item \textbf{Ecosystem-Wide Trends:} We identify trends across the AI agent ecosystem relating to systems' origin, role, level of agency, capabilities, safety, and transparency (\Cref{sec:findings}).
    \item \textbf{Case Studies:} We present three case studies of specific agents across three dominant interaction paradigms: a browser agent, an agentic chatbot, and a customizable enterprise agent builder (\Cref{sec:case_studies}).
\end{enumerate}

\section{Background and Related Work}\label{sec:background}
\textbf{Definitions of AI agents are nebulous and differ across fields.} The notion of artificial agency has a long and discordant history across disciplines, including cybernetics \citep{rosenblueth1943behavior, ashby1956introduction, wiener1961cybernetics}, artificial life \citep{maes1990designing, maes1993modeling, maes1995artificial}, rational agency \citep{rao1991modeling}, software engineering \citep{wooldridge1995intelligent, jennings2000agent}, reinforcement learning \citep{sutton2018reinforcement}, and philosophy \citep{dennett1989intentional, dung2024understanding}.
While definitions vary, they tend to emphasize related notions of autonomy, goal-directedness, and the ability to accomplish complex, long-horizon tasks.
Despite attempts to define the term ``agent'', including in the context of computational systems \citep{franklin1996agent, russell2020, kenton2023discovering, kasirzadeh2025characterizing}, \textit{we do \ul{not} decide among these definitions or offer an alternative}. Instead, we aim to synthesize elements of existing definitions related to a system's potential for economic and scientific impact (see \Cref{sec:inclusion_criteria}).

\textbf{The rise of AI agents:} \Cref{fig:gs_searches} illustrates the rapid increase in research focused on AI agents in recent years, particularly in 2025, with papers mentioning ``AI agent'' or ``agentic AI'' exceeding the total from 2020-–2024 combined by more than twofold. 
This has also been accompanied by a surge of interest in enterprise use of agents. 
For example, in a survey of 1,993 companies in June and July of 2025, McKinsey \& Company found that 62\% of respondents reported that their organizations were at least experimenting with AI agents \citep{Singla2025StateAI}.
Based on the estimated automatability of work across economic sectors, McKinsey also estimated that AI agents could automate 2.9 trillion dollars in US economic value by 2030.
Agents are also capable of automating increasing amounts of scientific research, having contributed to documented strides in life sciences, chemistry, materials science, physics, astronomy, and computer science \citep{gottweis2025towards, gridach2025agentic, wei2025ai, gibney2025ai, yamada2025ai}.   
As of 2025, AI agents have begun to write papers that have passed academic peer review \citep{sakanaai2025aiscientist}.
These estimates and reports are prone to conflicts of interest and hype \citep{kotliar2025can}, but they reflect an unmistakable rise in interest and prominence of AI agents.
Finally, as of 2026, recent MoltBook and OpenClaw Agents have arguably driven attention and concerns around AI agents to new heights \citep{manik2026openclaw, contreras2026openclaw, bastian2026malicious, gariuolo2026viral, cnbc2026openclaw}.

\textbf{Societal Risks and Ethical Concerns around AI Agents:} 
Just as AI agents enable unique opportunities, their ability to act in the real world in open-ended pursuit of goals presents new risks \citep{chan2023harms, cohen2024regulating, ruan2023identifying, gabriel2024ethics, bengio2025internationalaisafetyreport}.
For example, while chatbots often cause harm when human users act upon model outputs (e.g., deploying model-generated malicious code) \citep{phuong2024evaluating, kouremetis2025occult, mayoral2025cybersecurity}, agentic AI systems can \textit{directly} cause harm (e.g., autonomously hacking websites) \citep{fang2024llm, jaech2024openai, mayoral2025cybersecurity}.  
For these reasons, highly capable and agentic systems are often cited as a key risk factor for crises of accountability \citep{kolt2025governing, cooper2022accountability, himmelreich2019responsibility} and AI loss of control events \citep{bengio2023ai, hendrycks2023overview, bengio2025internationalaisafetyreport}.
Several prior works have focused on benchmarking agents' potential for specific harmful behaviors \citep{Andriushchenko2024AgentHarmAB, kumar2024refusal, US_AISI_2025, zou2025security, vijayvargiya2025openagentsafety}.
Meanwhile, others have argued that highly capable AI agents could contribute to systemic disruptions and risks, including to labor \citep{beraja2025generalized, deming2025technological, salari2025impacts}, inequality \citep{wang2025artificial, hammerschmidt2025bridging}, or the digital marketplace of ideas \citep{kazdan2024collapse, peterson2025ai, miklian2025new, ansari2025ai}.

\textbf{Mapping the AI Agent Landscape:}
This work follows the inaugural AI Agent Index from \citet{casper2025ai}.
Concurrently, the Princeton Holistic Agentic Leaderboard project \citep{kapoor2025holistic} curates evaluations of agentic AI systems across 9 benchmarks, and \citet{aiagentslist2025} maintains a list of over 600 ``agentic'' AI systems and products. 
Other works have studied agents by benchmarking their capabilities on economically valuable tasks \citep{patwardhan2025gdpval, vidgen2025ai, mazeika2025remote}, striving to increase visibility into their operation \citep{chan2024visibility, chan2024ids, chan2025infrastructureaiagents, yang2025survey, ning2025survey}, and studying their implications for economics and governance \citep{kasirzadeh2025characterizing, riedl2025ai, kolt2025governing, hadfield2025economy, kolt2026legal}.

\textbf{Documentation Frameworks:} Aiming to facilitate research and oversight \citep{winecoff2024improving}, a number of frameworks have been developed to document the features of AI systems, the resources used to build them, and the contexts in which they are deployed. These include datasheets \citep{gebru2018datasheets}, model cards \citep{mitchell2019model}, system cards \citep{gursoy2022system}, factsheets \citep{arnold2019factsheets}, AI nutrition facts \citep{twilio_nutrition_facts}, reward reports \citep{gilbert2022reward}, ecosystem graphs \citep{bommasani2023ecosystem}, data provenance cards \citep{longpre2023data}, eval cards \citep{dhar2025evalcards}, audit cards \citep{staufer2025audit}, usage cards \citep{wahle2023ai}, and safety cases \citep{clymer2024safety}.  
In addition, several databases have been created to collect information regarding contemporary AI systems and their real-world impacts, such as the Foundation Model Transparency Index \citep{bommasani2023foundation, bommasani20242024, wan20252025}, the AI Incident Database \citep{mcgregor2021preventing}, the AI Safety Index \citep{FLI2025}, and the AI Risk Repository \citep{slattery2024ai}. 
However, aside from the agent cards introduced here and in the inaugural AI Agent Index \citep{casper2025ai}, \textit{there are \ul{no} comparable frameworks for documenting agentic AI systems}.

\textbf{The 2024 AI Agent Index:} While this project follows the inaugural 2024 Index \citep{casper2025ai}, it represents a revision rather than a reiteration. To account for developments in the past year, this Index uses substantially revised inclusion criteria (\Cref{sec:inclusion_criteria}) and information fields (\Cref{sec:constructing}).
Most crucially, it indexes a \textit{smaller number of systems in greater depth} --- focusing on highly agentic systems with high-impact real-world applications. Unlike the 2024 Index, this Index also separates agentic chat interfaces, browser-based agents, and enterprise agent builders to reflect how popular agentic products are emerging in these three distinct forms (\Cref{sec:whats_included}). However, despite differences between last year's and this year's Indices, some trends are apparent. Both Indices' findings include distinct shortcomings of transparency related to safety and an increasing rate at which new qualifying systems are introduced (\Cref{sec:findings}).

\section{Constructing the 2025 AI Agent Index} \label{sec:constructing}

We constructed the 2025 AI Agent Index through systematic selection and annotation of deployed agentic systems. This section describes our inclusion criteria, emphasizing both agency and real-world impact, the scope of indexed systems, and our annotation methodology.

\begin{figure}
\centering
\Description{Flowchart of the index inclusion process. A box labeled candidate agent system flows left to right through Agency, Impact, and Practicality to a final box labeled included in index. Agency contains four required criteria: autonomy, goal complexity, environmental interaction, and generality. Impact contains three alternative criteria: public interest, market significance, and developer significance. Practicality contains three required criteria: public availability, deployability, and general purpose.}
\begin{tikzpicture}[
    node distance=0.3cm and 0.5cm,
    startbox/.style={rectangle, draw, minimum width=1.8cm, minimum height=1cm, align=center, font=\small, fill=blue!10},
    bigbox/.style={rectangle, draw, minimum width=2.8cm, minimum height=2.8cm, align=center, font=\small},
    critbox/.style={rectangle, draw, dashed, minimum width=2.2cm, minimum height=0.4cm, align=center, font=\scriptsize},
    endbox/.style={rectangle, draw, minimum width=1.8cm, minimum height=1cm, align=center, font=\small, fill=green!20},
    arrow/.style={-Stealth, thick},
    bigarrow/.style={-Stealth, thick}
]

\node[startbox] (candidate) {\textbf{Candidate}\\[-1pt]\textbf{Agent System}};

\node[bigbox, fill=gray!5, right=0.4cm of candidate] (agency) {};
\node[font=\small\bfseries, anchor=north] at (agency.north) [yshift=-0.1cm] {Agency};
\node[font=\tiny, anchor=north] at (agency.north) [yshift=-0.35cm] {(all required)};
\node[critbox, below=0.7cm of agency.north] (autonomy) {Autonomy};
\node[critbox, below=0.05cm of autonomy] (goal) {Goal complexity};
\node[critbox, below=0.05cm of goal] (environment) {Env. interaction};
\node[critbox, below=0.05cm of environment] (generality) {Generality};

\node[bigbox, fill=gray!5, right=0.4cm of agency] (impact) {};
\node[font=\small\bfseries, anchor=north] at (impact.north) [yshift=-0.1cm] {Impact};
\node[font=\tiny, anchor=north] at (impact.north) [yshift=-0.35cm] {(any required)};
\node[critbox, below=0.8cm of impact.north] (public) {Public interest};
\node[critbox, below=0.05cm of public] (market) {Market significance};
\node[critbox, below=0.05cm of market] (developer) {Developer significance};

\node[bigbox, fill=gray!5, right=0.4cm of impact] (practicality) {};
\node[font=\small\bfseries, anchor=north] at (practicality.north) [yshift=-0.1cm] {Practicality};
\node[font=\tiny, anchor=north] at (practicality.north) [yshift=-0.35cm] {(all required)};
\node[critbox, below=0.8cm of practicality.north] (availability) {Public availability};
\node[critbox, below=0.05cm of availability] (deployability) {Deployability};
\node[critbox, below=0.05cm of deployability] (general) {General purpose};

\node[endbox, right=0.4cm of practicality] (included) {\textbf{Included in}\\[-1pt]\textbf{Index}};

\draw[bigarrow] (candidate) -- (agency);
\draw[bigarrow] (agency) -- (impact);
\draw[bigarrow] (impact) -- (practicality);
\draw[bigarrow] (practicality) -- (included);

\end{tikzpicture}
\caption{Inclusion criteria for the Index. Candidate agents flow through three criteria categories from left to right. Systems must satisfy all agency criteria, at least one impact criterion, and all practicality criteria. See \Cref{sec:inclusion_criteria} for details of each criterion.}
\label{fig:inclusion_criteria}
\end{figure}
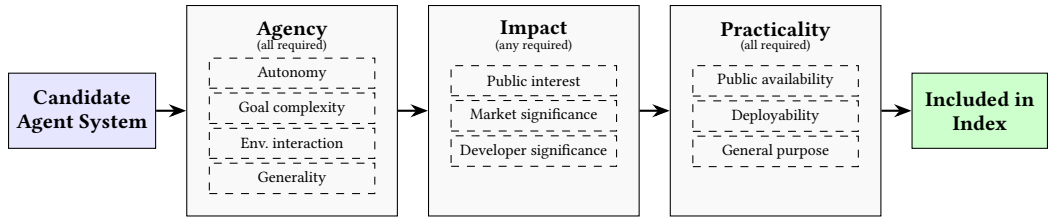

\subsection{Inclusion Criteria for Agents} \label{sec:inclusion_criteria}

To determine whether a system is included in the Index, we use a set of criteria for a system's \textit{agency}, its \textit{impact}, and its \textit{practicality} to index. To be included, systems must satisfy \textit{all} agency criteria, \textit{at least one} impact criterion, and \textit{all} practicality criteria. All criteria were evaluated as of the Index's cutoff date of December 31, 2025. 

\textbf{Agency criteria (\ul{all} required for inclusion).} Rather than proposing a new definition of agency, we draw on prior literature and follow the approaches developed by \citet{chan2023harms}, \citet{kasirzadeh2025characterizing}, and \citet{fengLevelsAutonomyAI2025}, which characterize AI agents as systems that exhibit, to some significant degree, a combination of the following properties. For our ``agency'' criterion to be met, all four of the following must be satisfied:

\begin{enumerate}
  \item \textbf{Autonomy.} Included agents must be able to operate with minimal human oversight and make consequential decisions without continuous user input \citep{chan2023harms,kasirzadeh2025characterizing}. \citet{fengLevelsAutonomyAI2025} conceptualize autonomy as a spectrum characterized by the user's role: operator, collaborator, consultant, approver, or observer. We require at least intermediate autonomy: ``the AI system can perform the majority of tasks independently, though it still relies upon input from the principal for critical determinations'' \citep{kasirzadeh2025characterizing}. This corresponds to autonomy Level 2 (L2): ``user and agent collaboratively plan, delegate, and execute'' from \citet{fengLevelsAutonomyAI2025}.

  \item \textbf{Goal complexity.} Included agents must be able to pursue high-level objectives (e.g., ``make money'') through long-term planning, breaking down complex goals into subgoals, and making temporally dependent decisions \citep{chan2023harms,kasirzadeh2025characterizing}. In practice, we operationalize this as an agent being reliably capable of at least three autonomous tool calls and high-level task specification without step-by-step instructions.

  \item \textbf{Environmental interaction.} Included agents must be able to directly interact with the world through tools and APIs, creating substantial changes in their environment \citep{chan2023harms,kasirzadeh2025characterizing}, rather than merely conversing with users. In practice, this requires write access to a computer and the ability to choose tools.

  \item \textbf{Generality.} Included agents must be able to handle under-specified instructions and adapt to new tasks, demonstrating versatility across related tasks rather than single narrow functions \citep{chan2023harms,kasirzadeh2025characterizing}.
\end{enumerate}

\textbf{Impact criteria (\ul{any} required for inclusion).} To focus on agents with significant real-world influence, at least one of the following must be satisfied:

\begin{enumerate}
  \item \textbf{Public interest.} Substantial search volume of at least 10,000 searches or GitHub stars for open-source projects of at least 20,000 in total.\footnote{This uses Google search number estimates across the top five keywords for 2025. We use the ``historical\_volume'' field of the \href{https://docs.ahrefs.com/docs/api/keywords-explorer/schemas/overview}{Ahrefs API} as the data source. Limitation: Agents embedded in broader products may not be searched by their specific agent name. See \Cref{app:search} for mitigations. Enterprise agents typically have lower search volume than end-user products.}

  \item \textbf{Market significance.} The developer has a market capitalization or valuation $\geq$ \$1 billion USD. To determine this, we collected data from stock exchanges, Crunchbase, and Epoch AI.

  \item \textbf{Developer significance.} The developer is a member of the 2024 Foundation Model Transparency Index \citep{bommasani20242024}, Frontier Model Forum \citep{frontiermodelforumMembership}, or a signatory of the Frontier AI Safety Commitments \citep{FrontierAISafety2025} or Artificial Intelligence Safety Commitments \citep{chinaacademyofinformationandcommunicationstechnologyFirst17Companies2024}.
\end{enumerate}

\textbf{Practicality (\ul{all} required for inclusion).} To ensure analysis reflects deployed systems accessible for evaluation, all three of the following criteria must be satisfied.

\begin{enumerate}
  \item \textbf{Public availability.} Included agents must be publicly accessible products. This excludes company-internal products or limited pre-releases. We determined this based only on publicly available information, such as blog posts, documents, or demos.
  
  \item \textbf{Deployability.} Included agents must be able to perform tasks off the shelf with minimal configuration and no software engineering expertise. This distinguishes ready-to-use agents from development frameworks.

  \item \textbf{General purpose.} Included agents must be capable of performing general-purpose tasks in practice, regardless of how they are advertised. This excludes domain-specific agents (e.g., coding-only or legal analysis agents). Claude Code and similar tools, though advertised as coding agents, are included insofar as they can perform general-purpose tasks \textit{through code}. This criterion is included to reduce the scope to those agents with the broadest impact.
\end{enumerate}

\subsection{What Does the Index Include?}\label{sec:whats_included}

We identify three distinct types of agents, each with different interfaces. We divide agents into these three categories based on how users primarily interact with and operate them.\footnote{These categories are not generally exhaustive but represent the common interaction types across the 30 identified agents.} These different modalities present distinct technical architectures and governance challenges.

\begin{itemize}
    \item \textbf{\ulchat{Chat applications with agentic tools (12 systems).}} This category primarily includes chat interfaces with extensive tool access. This includes general-purpose coding agents (Claude Code) that operate through terminal interfaces with broad capabilities, but excludes narrow coding-only agents (GitHub Copilot). \textbf{Examples}: Manus AI, ChatGPT Agent, Claude Code.
    \item \textbf{\ulbrowser{Browser-based agents (5 systems).}} These are agents whose primary interface is browser or computer use, with extensive browser/computer interaction tools. They are distinct from chat agents with web search capabilities (ChatGPT web search, Claude web search), which primarily perform retrieval and summarization. Browser-based agents present higher risks through background execution, event triggers, and direct transactions. We also include system-based agents that run directly on mobile or desktop devices in this category. \textbf{Examples}: Perplexity Comet, ChatGPT Atlas, ByteDance Agent TARS.
    \item \textbf{\ulent{Enterprise workflow agents (13 systems).}} These are business management platforms with agentic features aimed at reliably automating business tasks. Typically implemented as workflow builders with agentic actions within nodes. \textbf{Examples}: Microsoft Copilot Studio, ServiceNow Agent.
\end{itemize}

\subsection{How Were Agents Identified?}

LLM-based research queries surfaced 95 candidate agents (see \Cref{app:prompts_deep_research} for details). These were screened against our inclusion criteria. Ambiguous cases were included for in-depth annotation, with final inclusion decisions made after full evaluation. We consulted two Chinese ecosystem experts to mitigate linguistic or ecosystem-related blind spots. We also cross-referenced our list of candidate agents against the 2024 Index \citep{casper2025ai}, the Princeton Holistic Agent Leaderboard \citep{kapoor2025holistic}, and \citet{aiagentslist2025}.
Finally, recognizing the possibility that we may have missed an agent that meets our inclusion criteria, we have established a structured process for facilitating further corrections to the Index.
These can be submitted at \url{https://aiagentindex.mit.edu/feedback}.

For companies offering both off-the-shelf agents and custom agent builders targeting comparable use cases, we combined these into a single listing and documented the most capable agents that users could create or deploy through either offering. We did not combine offerings when they targeted different audiences (e.g., consumer-facing chat agents versus enterprise agent builders).

\subsection{How Were Agents Annotated?}

\textbf{We annotated agents across six categories}: \ul{product overview} (release date, pricing, description), \ul{company \& accountability} (developer entity, governance documents, contact mechanisms), \ul{technical capabilities} (models, tools, architecture, memory), \ul{autonomy \& control} (autonomy levels, approval requirements, monitoring, emergency stops), \ul{ecosystem interaction} (identification protocols, interoperability standards, web conduct), \ul{safety \& evaluation} (guardrails, sandboxing, evaluations, third-party testing, compliance). This resulted in a total of 45 fields of information per system. See \Cref{app:annotation_fields} for a full list of all 45. We also include the \ul{inclusion criteria} (search volume, market capitalization). These categories expanded upon the 2024 Index \citep{casper2025ai} and were revised through discussion with subject-matter experts. See \Cref{app:changes_annotation_fields} for a full account of this year's fields compared to the 2024 Index's.

\textbf{We annotated only public information} from documentation, websites, demos, published papers, and governance documents. We did not perform experimental testing (e.g., probing agent behavior or running benchmarks). See \Cref{app:index} for the full list of sources used. All web sources linked in the Index were archived. When possible, we created accounts and used demos to explore agent interfaces directly.

\textbf{Seven of the paper's authors annotated agents according to their domain expertise}. To ensure consistency, experts were each responsible for categories of fields rather than specific agents. 
Annotations emphasized object-level findings over interpretations and focused exclusively on agent-specific rather than underlying model features. For platforms creating agents, annotations assessed the most capable version of each agent that could be readily configured, documenting capabilities, limitations, and default configurations. ``None found'' indicates we found no public information; ``None'' indicates confirmed absence; ``Not applicable'' indicates irrelevance of field to this agent.

\textbf{Annotations followed detailed protocols developed iteratively through calibration exercises}; see \Cref{app:annotation_guide}. Inter-annotator consistency was maintained through protocol revisions and cross-validation. All annotations were independently reviewed by at least one other annotator. 37 out of 1,350 fields with discrepancies were resolved through discussion. Finally, we used GPT-5.2 with web search to screen annotations for potential inaccuracies; see \Cref{app:verify}.

\textbf{Companies were contacted and given four weeks to correct annotations}. 23\% offered some form of response at the time of publication, but only 4/30 provided substantive comments.\footnote{These response rates were lower than for the 2024 Index \citep{casper2025ai}, which we attribute to how the 2024 Index used broader inclusion criteria, which included a number of agents created by academic research groups who had a high response rate.}
Their comments have been incorporated into the final Index. An ongoing correction form remains available for updates via \url{https://aiagentindex.mit.edu/feedback}.

\section{Findings} \label{sec:findings}

\begin{figure}
    \centering
    \includegraphics[width=1\linewidth]{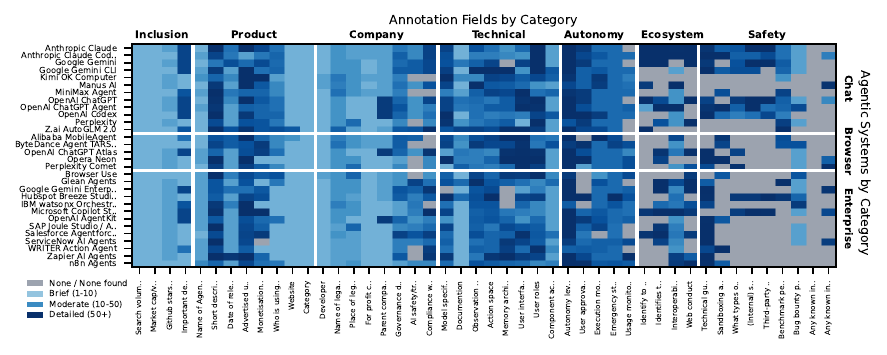}
    \Description{Heatmap of annotation detail by field and agent. Rows are agents and columns are fields grouped into inclusion, product, company, technical, autonomy, ecosystem, and safety. The color scale runs from gray for none or none found to darker blue for more detailed entries. Most product, company, and technical fields are filled, while ecosystem and especially safety contain many gray cells across many agents.}
    \caption{For 227 out of 1,350 fields, we were unable to find any information (gray). This is most common in the ``Ecosystem Interaction'' and ``Safety, Evaluation, and Impact'' categories. Non-empty information fields are 14 words long on average. See \Cref{app:transparency_gaps_full} for a full page version.}
    \label{fig:transparency_gaps}
\end{figure}

We present findings from the 2025 AI Agent Index across six categories: product overview, company and accountability, technical capabilities and system architecture, autonomy and control, ecosystem interaction, and safety, evaluation, and impact. \Cref{fig:transparency_gaps} shows the full Index with annotations for all 30 agents. We reveal patterns in how agents are deployed, governed, and documented, alongside significant transparency gaps around safety, evaluation practices, and ecosystem interaction. See \Cref{app:index} for details on accessing the full Index.

\subsection{Product Overview}

\textbf{Most agents were released in 2024--2025, indicating a recent surge in agent deployment.} 24/30 agents were released or received major agentic feature updates during this period, with earlier systems like ChatGPT (2022) and Perplexity (2022) adding agentic capabilities later. While the underlying models (such as GPT-4) are older, agents meeting our inclusion criteria are emerging at an increasing rate, with a surge of releases in late 2024 and 2025 (see \Cref{fig:agent_release_timeline,fig:agent_releases_per_month}). This separates capability (frontier models) from productization (agentic scaffolding).

\textbf{Chat interfaces are the most abundant, followed closely by enterprise workflow platforms.} 12/30 agents use conversational chat interfaces, 13/30 are enterprise automation platforms, and 5/30 are browser-based agents focused on Graphical User Interface (GUI) operation. Notably, Chinese GUI agents are more commonly designed with phone-use and computer-use capabilities (3/5). 

\textbf{Advertised use cases cluster around three themes that cut across agent categories.} Research and information synthesis appears in 12/30 agents spanning both consumer chat assistants and enterprise platforms. Workflow automation across business functions (HR, sales, support, IT) is advertised by 11/30 agents, concentrated in enterprise products. GUI/browser operation for tasks like forms, ordering, and booking is emphasized by 7/30 agents, primarily browser-based agents, but also appearing in some chat products.

\subsection{Company and Accountability}

\begin{figure}
    \centering
    \begin{subfigure}[b]{0.45\linewidth}
        \centering
        \includegraphics[width=\linewidth]{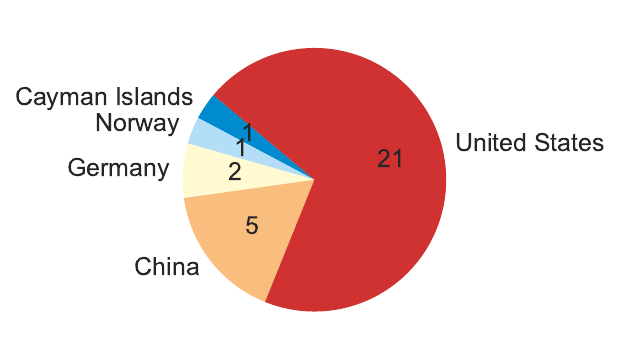}
        \Description{Pie chart of company incorporation. The United States is the dominant slice with 21 of 30 companies. China is next with 5, Germany has 2, and Norway and the Cayman Islands have 1 each.}
        \caption{The majority of companies are incorporated in the US (21/30), followed by China (5/30).}
        \label{fig:country_distribution}
    \end{subfigure}
    \hfill
    \begin{subfigure}[b]{0.45\linewidth}
        \centering
        \includegraphics[width=0.75\linewidth]{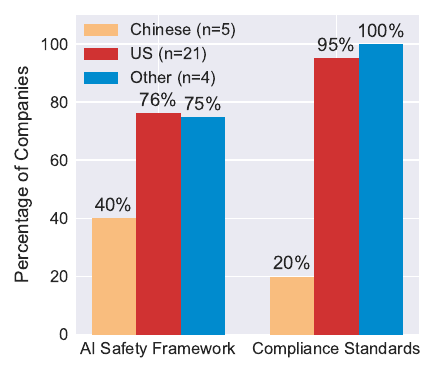}
        \Description{Grouped bar chart of two disclosure measures by region. The horizontal axis has AI safety framework and compliance standards; the vertical axis shows percentage of companies. Chinese companies are lowest on both measures at 40 percent and 20 percent, compared with 95 and 100 percent for US companies and 76 and 75 percent for other companies.}
        \caption{Chinese companies, except Z.ai, do not publish AI safety and trust frameworks or standards adherence.}
        \label{fig:safety_country_distribution}
    \end{subfigure}
    \Description{Two-panel figure on developer geography and disclosure. The left panel is a pie chart showing that most companies are incorporated in the United States, with China a distant second. The right panel is a grouped bar chart showing lower publication of safety frameworks and compliance standards among Chinese companies than among US or other companies.}
    \caption{Comparison between Chinese, US, and other agent developers. To mitigate potential blind spots, two native Chinese speakers reviewed our coverage of safety frameworks, including those published only in Mandarin. Documentation in other languages may not be fully captured; see \Cref{sec:limitations}.}
    \label{fig:combined}
\end{figure}

\textbf{Agent developers are concentrated in the United States and China, with limited representation from other regions.} 13/30 agents are developed by Delaware-incorporated companies spanning both large incumbents and startups. 5/30 agents are China-incorporated. Non-US, non-China incorporations are less common (4/30), including Germany (SAP, n8n), Norway (Opera), and Cayman Islands (Manus); see \Cref{fig:country_distribution}.

\textbf{Chinese agents represent a distinct geographical cluster with different governance patterns.} China-incorporated agents typically lack the safety frameworks (1/5) and compliance standards (1/5) common among other agents. However, their compliance may simply not be documented publicly. See \Cref{fig:safety_country_distribution} for a comparison.

\textbf{Only half of agent developers publish safety or trust frameworks, though enterprise assurance standards are more common.} 15/30 agents reference AI safety frameworks like Anthropic's Responsible Scaling Policy, OpenAI's Preparedness Framework, or Microsoft's Responsible AI Standard \citep{anthropicResponsibleScalingPolicy2025,openaiPreparednessFramework2025,microsoftResponsibleAIStandard2022}. 10/30 agents have no safety framework documentation. Enterprise assurance standards (\href{https://www.aicpa-cima.com/topic/audit-assurance/audit-and-assurance-greater-than-soc-2}{SOC 2}, \href{https://www.iso.org/standard/27001}{ISO 27001}, \href{https://www.fedramp.gov/}{FedRAMP High}, \href{https://www.iso.org/standard/42001}{ISO/IEC 42001}) are more widely adopted. 5/30 agents have no compliance standards documented.

\subsection{Technical Capabilities and System Architecture}

\textbf{Most agents use a small set of closed-source frontier models as their backend.} Only frontier AI companies themselves (Anthropic, Google, OpenAI) and Chinese developers run their own proprietary models; the majority rely primarily on GPT, Claude, or Gemini model families. Enterprise agents are typically model-agnostic, with 9/30 agents explicitly supporting user selection across providers; see \Cref{fig:technical_by_category} for a comparison across agent categories.

\textbf{Action spaces differ systematically by agent category, with widespread Model Context Protocol (MCP) support but varied implementation.} Enterprise workflow agents act via Customer Relationship Management (CRM) connectors and record updates (8/13), Command Line Interface (CLI) agents use filesystem edits and terminal commands (4/30), and browser agents manipulate web pages through click/type/navigate actions (5/5). 20/30 agents support Model Context Protocol (MCP) for tool integration, though proprietary connectors are often promoted over open MCP servers. Enterprise agents tend to have more constrained action spaces and prioritize guardrails around tool use. See \Cref{fig:technical_by_category} for a comparison of technical features by agent category.

\textbf{Most qualifying agents are closed source at the product level despite growth in the open agent ecosystem.} 23/30 agents are fully closed. 7/30 agents open-source their agent framework or harness (Alibaba MobileAgent, Browser Use, ByteDance Agent TARS, Google Gemini CLI, n8n Agents, OpenAI Codex, WRITER).

\textbf{Canvas-based user interfaces are the standard for agent design workflows.} 8/13 enterprise platforms use visual composition interfaces (Glean, Google Gemini Enterprise, HubSpot Breeze, n8n, Microsoft Copilot Studio, OpenAI AgentKit, Salesforce Agentforce, Zapier) for building agents, while chat interfaces dominate end-user operation (14/30 agents).

\subsection{Autonomy and Control}

\begin{figure}
    \centering
    \includegraphics[width=0.75\linewidth]{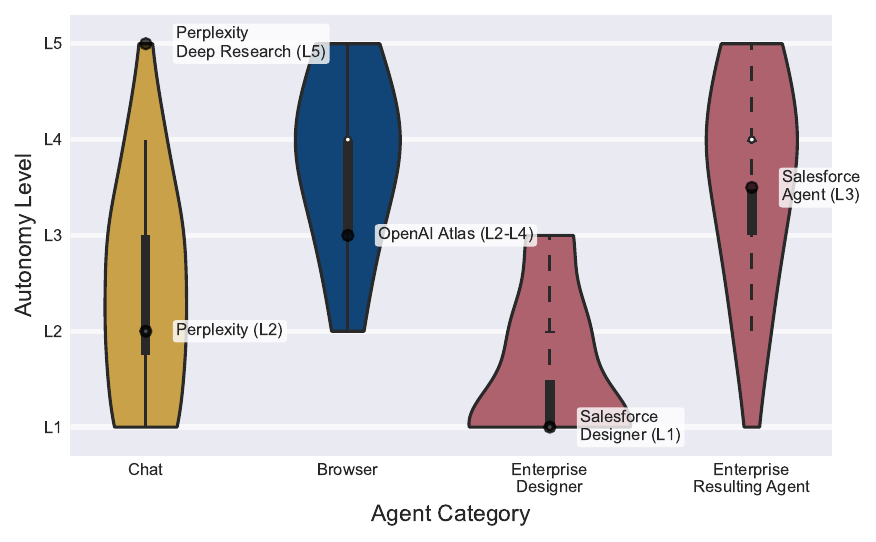}
    \Description{Violin plots of autonomy levels by agent category. The vertical axis runs from level 1 to level 5. Chat agents cluster at lower levels with a long tail to level 5; browser agents cluster between levels 3 and 5; enterprise is split into designer agents near level 1 and resulting agents near levels 3 to 4. Three labeled points mark Perplexity, OpenAI Atlas, and Salesforce Agent.}
    \caption{Distribution of levels of autonomy across each agent category, with the autonomy of three representative agents marked. For each category, certain levels of autonomy are more common (shown as wider). \ulbrowser{Browser-based} and \ulent{deployed enterprise agents} are the most agentic (L5). The resulting agents deployed through enterprise designers (e.g. L3) are significantly more agentic than the process of designing the agents (e.g. L1).}
    \label{fig:distribution_autonomy_levels}
\end{figure}

\textbf{\ulchat{Chat-first assistants maintain lower autonomy} (L1-L3) with turn-based interaction.}\footnote{Following \citet{fengLevelsAutonomyAI2025} we conceptualize autonomy as a spectrum characterized by the user's role from L1 (user directs and makes decisions) to L5 (agent operates with
full autonomy and user observes). Most agents operate across a range of levels. More autonomy is not necessarily better.} Anthropic Claude, Google Gemini, and OpenAI ChatGPT operate in a turn-based paradigm where the agent executes a single set of actions and waits for the next user prompt (3/30). Within a single product, autonomy can vary substantially. For example, ``regular chat'' (L1) versus ``deep research'' (L3-L5). See \Cref{fig:distribution_autonomy_levels} for the spectrum of autonomy levels for each agent category.

\textbf{\ulbrowser{Browser agents operate with significantly higher autonomy} (L4-L5), offering limited opportunities for mid-execution intervention.} Browser Use's agent and Perplexity's Comet perform tasks autonomously once prompted, with no means for user involvement during execution. Once a query is sent, users cannot easily intervene or steer the agent until it finishes (2/5 browser agents, 5/30 overall).

\textbf{\ulent{Enterprise platforms show a design/deployment autonomy split.}} During the design phase, users manually configure triggers, actions, and guardrails using visual canvases. 8/13 platforms provide AI assistance with the design process itself (L1-L2).
Once deployed, these agents often operate at L3-L5 autonomy, triggered by events like a new email or a database change, without any human involvement during the actual task execution (6/30 agents: Glean, Google Gemini Enterprise, IBM watsonx, Microsoft Copilot Studio, n8n, OpenAI AgentKit).

\textbf{User approval mechanisms are implemented selectively based on the task's risk level, with some agents offering live oversight modes.} Developer/Command-Line-Interface (CLI) agents require explicit confirmations for sensitive operations like file edits and command execution (3/30), while browser agents gate only high-risk steps like authentication and payments (4/30). Some agents offer ``watch mode'' for real-time oversight of critical actions (5/30 agents, including ChatGPT Agent/Atlas, Opera Neon). 

\textbf{Execution traces and monitoring are common, but the scope and level of transparency vary widely.} 10/30 agents provide detailed action traces with visible chain-of-thought reasoning. 6/30 agents show summarized reasoning without detailed tool traces. For many enterprise agents, it is unclear from publicly available information whether monitoring for individual executions exists. 12/30 agents provide no usage monitoring or only notices once users reach the rate limit. 

\textbf{Most agents allow user-initiated stopping, but some lack fine-grained stop controls.} 20/30 agents document pause/stop mechanisms. 4/30 agents (Alibaba MobileAgent, HubSpot Breeze, IBM watsonx, n8n) lack documented stop options despite autonomous execution. For enterprise platforms, there is sometimes only the option to stop all agents or retract deployment.

\subsection{Ecosystem Interaction}

\textbf{Model Context Protocol (MCP) is the dominant interoperability standard across the ecosystem}, supported by 20/30 agents. The Agent-to-Agent (A2A) protocol is supported by 6/30 agents, all of which are enterprise platforms.

\textbf{Most agents do not disclose their AI nature to end users or third parties by default.} 21/30 agents have no documented default disclosure behavior. Only 3/30 agents support watermarking generated media (e.g., through SynthID and C2PA \citep{dathathriScalableWatermarkingIdentifying2024, C2PA}). Enterprise platforms generally shift the burden of disclosure to the customer, meaning that the obligation to inform end users that they are interacting with an AI system falls on the business deploying the agent.

\textbf{Technical identification varies widely, with many agents blending into normal traffic.} Only 7/30 agents publish stable User-Agent (UA) strings and IP address ranges for verification. 6/30 agents explicitly use Chrome-like UA strings and residential/local IP contexts, mimicking human web traffic. OpenAI ChatGPT Agent is unique in providing cryptographic request signing via HTTP Message Signatures (RFC 9421 \citep{spornyHTTPMessageSignatures2024}), in which each outbound HTTP request carries a digital signature that receiving websites can verify against a published public key to confirm the request originated from ChatGPT Agent.

\textbf{Robots.txt compliance varies by agent interaction type.} 6/30 agents explicitly state that their crawler bots respect robots.txt. However, agents designed to execute tasks on behalf of users often ignore standard exclusion protocols. For example, BrowserUse's agent markets bypassing anti-bot systems and browsing ``like a human.''

\textbf{Web conduct practices are undocumented for most agents.} 16/30 agents provide no clear statement about robots.txt, CAPTCHA handling, or web access methods, particularly for enterprise platforms where web access occurs via third-party scrapers or search API connectors.

\subsection{Safety, Evaluation, and Impact}

\textbf{Technical guardrails against potentially harmful actions differ systematically by agent type, with consumer agents providing built-in guardrails while builder platforms delegate to users.} Consumer-facing agents typically limit the permissions and action space of tools (8/30 agents) and provide defenses against prompt-injection attacks (7/30). Agent builder platforms (Zapier, Salesforce, OpenAI AgentKit) generally provide optional guardrail modules and little information on built-in protections. Sandboxing or VM isolation is documented for 9/30 agents, primarily developer/CLI tools and browser agents. 9/30 agents have no guardrails documented, and 7/13 enterprise agents describe options for setting up guardrails, but no sandboxing or containment.

\begin{figure}[t]
    \centering
    \begin{subfigure}[t]{0.54\textwidth}
        \vspace{0pt}
        \centering
        \includegraphics[width=\linewidth]{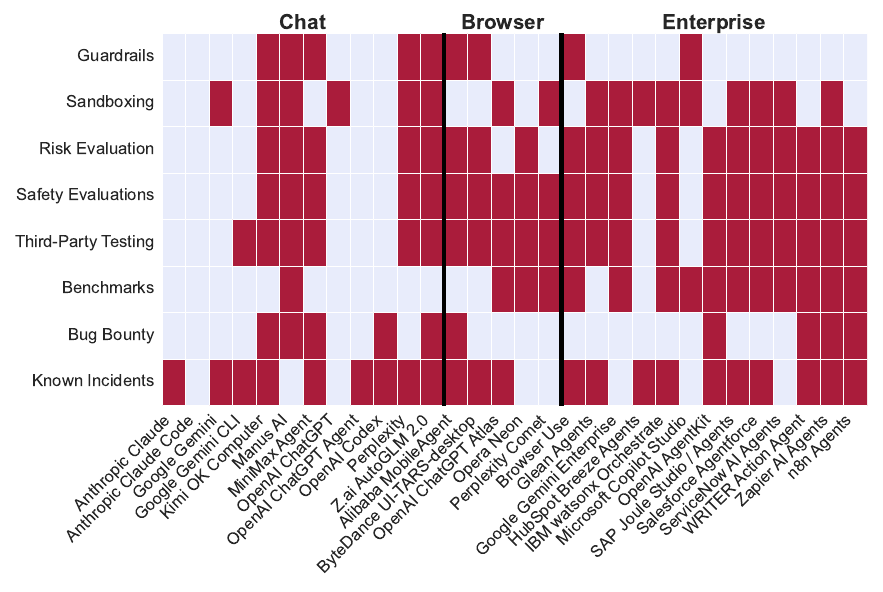}
        \Description{Matrix of missing safety-related information by agent. Columns are agents grouped into chat, browser, and enterprise; rows are guardrails, sandboxing, risk evaluation, safety evaluations, third-party testing, benchmarks, bug bounty, and known incidents. Red cells mark fields with no information found. Red cells appear across all categories and are especially dense for browser and enterprise agents.}
        \caption{Safety fields with no public information are highlighted in red.}
        \label{fig:safety-none-found}
    \end{subfigure}
    \hfill
    \begin{subfigure}[t]{0.45\textwidth}
        \vspace{0pt}
        \centering
        \includegraphics[width=\linewidth]{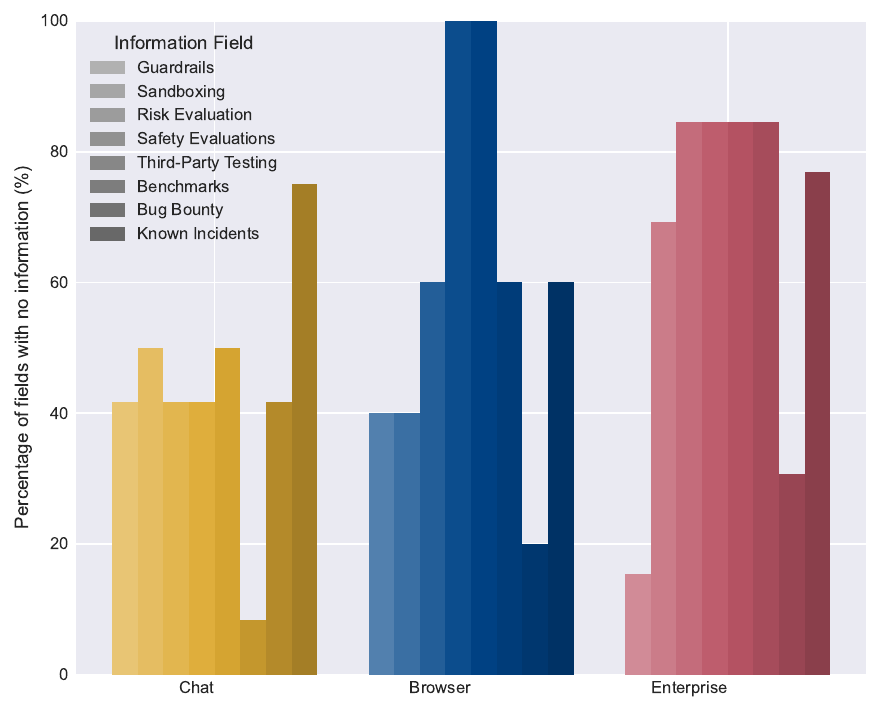}
        \Description{Grouped bar chart of missing safety information by category. The vertical axis shows percentage of fields with no information. Browser agents have the highest missingness overall, including 100 percent for risk evaluation and safety evaluations, while enterprise agents are also high on several fields and chat agents are lower but still substantial.}
        \caption{Percentage of \textit{missing} information by agent category.} 
        \label{fig:safety-by-category}
    \end{subfigure}
    \Description{Two-panel figure on missing safety information. The left panel shows a red cell matrix of missing fields for each agent. The right panel summarizes the same fields as percentages by chat, browser, and enterprise categories. Both panels show browser and enterprise agents with more missing safety information than chat agents.}
    \caption{Most safety, evaluation, and social impact related fields (135/240) have no information available. Enterprise agents (69/104, 66\%) and browser agents (24/40, 60\%) have the most missing fields, followed by chat agents (42/96, 44\%).}
    \label{fig:safety-comparison}
\end{figure}

\textbf{Safety evaluation practices diverge sharply between frontier AI companies and enterprise platforms, with most agents providing no evaluation information.} Frontier AI companies (OpenAI, Anthropic, Google) focus on existential and behavioral alignment risks, publishing system cards covering catastrophic risks, autonomy, and misuse (7/30 agents). Enterprise platforms define safety primarily through compliance and data security rather than agent-specific evaluations. Few evaluations test the agentic setup rather than base model components; only ChatGPT Agent, OpenAI Codex, Claude Code, and Gemini 2.5 Computer Use have agent-specific system cards. 25/30 agents disclose no internal safety results, and 23/30 agents have no third-party testing information. Third-party testing is documented for only 3/30 agents (Anthropic Claude, OpenAI ChatGPT, OpenAI Codex). 18/30 agents operate bug bounties or vulnerability disclosure programs.

\textbf{Large frontier AI companies lead on thorough safety reporting.} The most consistent and extensive reporting on safety came from large AI companies (OpenAI, Google, Anthropic, Microsoft). Anthropic's Claude Code was the only system we studied for which we found information on all 8 safety fields. In contrast, Moonshot AI and Manus were tied for providing the least safety-relevant information, with only 1 of 8 safety fields available. 

\textbf{Documented security incidents concentrate in browser agents and relate to prompt injection.} 8/30 agents have known incidents or reported security concerns (Anthropic Claude Code, Google Gemini Enterprise, Manus AI, Microsoft Copilot Studio, OpenAI ChatGPT, Opera Neon, Perplexity Comet, ServiceNow AI Agents). Prompt injection vulnerabilities are documented for 2/5 browser agents. 

\textbf{A transparency gap exists between capability benchmarks and safety documentation.} 9/30 agents report capability benchmarks (GUI/computer-use or coding), but the same agents often lack safety evaluation disclosure.

\section{Illustrative Case Studies} \label{sec:case_studies}

The Index annotations provide a structured view across the agent ecosystem, but examining individual agents in depth reveals how different categories operationalize autonomy, safety, and accountability in practice. We present three case studies representing distinct agent categories: ChatGPT Agent (\ulchat{agentic chat interface}), Perplexity Comet (\ulbrowser{browser-based agent}), and HubSpot Breeze Agents (\ulent{enterprise agent builder}). 

We selected these agents based on public interest within their respective categories, while avoiding featuring the same company twice. Each agent exemplifies characteristic features of its category while raising distinct questions about ecosystem dependencies, safety practices, and transparency. These case studies are not endorsements or critiques of the specific systems; rather, they serve as reference points for understanding how agents from different categories manifest common and divergent patterns in the Index.

\subsection{Chats With Agentic Tools: ChatGPT Agent}

\begin{outlinebox}[chat_yellow]

ChatGPT Agent operates within its chat interface but can interact with websites directly on users' behalf. The system demonstrates L2-L4 autonomy, following \citet{fengLevelsAutonomyAI2025}. User approval is required for sensitive operations like checkout. The agent operates in a hosted virtual computer with actions performed in a virtual browser and sandbox terminal with limited network access.

ChatGPT Agent is one of two systems in the Index (alongside Gemini 2.5 Computer Use) providing a dedicated agent-specific system card. This contrasts with most chat-based agents, which rely on base model documentation. OpenAI evaluates ChatGPT Agent across usage policy compliance, jailbreaks, hallucinations, fairness, CBRN risks, cyber capabilities, and autonomy using established benchmarks. The system is the only one implementing cryptographic signing of HTTP requests (\citet{openai_agent_allowlisting}; RFC 9421 \citep{spornyHTTPMessageSignatures2024}), addressing identity and auditability challenges that affect all browser agents, including Perplexity Comet and Opera Neon.

\textit{A governance challenge for chat agents is that a single interface spans from passive Q\&A (L1) to autonomous web actions (L4), meaning users may not anticipate when a request triggers consequential real-world actions. In principle, vertical integration from model to deployment enables ChatGPT Agent to coordinate safety across this spectrum, but this approach is unavailable to the majority of chat agents that depend on third-party foundation models.}

\end{outlinebox}

\subsection{Browser-based agents: Perplexity Comet}

\begin{outlinebox}[browser_blue]
Perplexity Comet operates at L4-L5 autonomy, the highest in the Index and representative of other browser agents that execute autonomously rather than through turn-based interaction. Unlike ChatGPT Agent's approval gates for sensitive operations, Comet proceeds autonomously once initiated. We found no agent-specific safety evaluations, third-party testing, or benchmark performance disclosures. Perplexity has published research on prompt injection mitigation \citep{zhangBrowseSafeUnderstandingPreventing2025}, but has not documented safety evaluation methodology or results for Comet. No sandboxing or containment approaches beyond prompt-injection mitigations were documented.

Security researchers identified multiple prompt injection vulnerabilities in 2025, including indirect injection, where malicious webpage content could be executed as commands, and URL-based attacks extracting data from connected services \citep{chaikinAgenticBrowserSecurity2025,gispanCometJackingHowOne2025}. These incidents illustrate fundamental challenges for browser agents processing untrusted web content, similar to vulnerabilities found in Opera Neon.

Perplexity argues that AI assistants ``work just like a human assistant'' when fetching content on behalf of users to justify user-driven agents ignoring robots.txt restrictions \citep{perplexityteamAgentsBotsMaking2025}. Perplexity publishes user-agent strings for \lstinline|Perplexity Bot| and \lstinline|Perplexity-User|, but Cloudflare documented undeclared crawlers using generic signatures to evade blocks \citep{corralPerplexityUsingStealth2025}. Amazon threatened legal action over Comet not identifying itself as an agent \citep{sriramAmazonSuesPerplexity2025}.

\textit{Browser agents face the challenge of operating in environments controlled by third parties who may have no relationship with the agent developer and no mechanism to negotiate or verify terms of interaction, as existing web protocols like robots.txt were designed for crawlers, not autonomous actors. While this lack of established trust relationships presents risks, zero-trust interaction can also lead to more robust security standards and protocols. Comet combines the highest autonomy in the Index with minimal safety disclosure and no documented third-party testing.}
\end{outlinebox}

\subsection{Enterprise Agent Builders: HubSpot Breeze Agents}

\begin{outlinebox}[enterprise_red]
HubSpot Breeze enables organizations to create workflow agents through templated configurations. Users fill fields in base prompt templates rather than coding, a low-code approach shared across enterprise builders in the Index. Breeze demonstrates split autonomy common to this category: L1-L3 during design, L5 when deployed with automatic triggers based on data changes or events.

Users configure whether actions require approval during creation (required by default), though automatically triggered agents can operate without approval in background workflows. Breeze's action space focuses on internal databases and organizational tools, creating a natural sandbox that limits external impact. Behavior constraints operate through tool permissions and user roles rather than content-level guardrails.

Safety disclosure follows a compliance-focused pattern typical of enterprise platforms. Breeze uses PurpleLlama as a model protection layer and underwent penetration testing by PacketLabs, but provides no methodology, results, or testing entity details. The platform maintains compliance certifications (e.g., \href{https://www.aicpa-cima.com/topic/audit-assurance/audit-and-assurance-greater-than-soc-2}{SOC 2}, \href{https://eur-lex.europa.eu/eli/reg/2016/679/oj}{GDPR}, \href{https://www.hhs.gov/hipaa/index.html}{HIPAA}) and trust center documentation. Model selection occurs automatically, supporting only OpenAI models.

\textit{For enterprise builders, the deployed agent is a joint product of the platform and the business user, but neither fully controls or has visibility into the other's contribution. The platform cannot anticipate all deployment contexts; the deployer cannot inspect underlying model behavior. Again, this information asymmetry can lead to more robust solutions. Breeze Agents emphasize compliance, while agent-specific guardrails become the user's responsibility.}
\end{outlinebox}

\section{Discussion}

The 2025 AI Agent Index provides information across 1,350 fields for 30 prominent AI agents. Beyond the findings detailed above, we identify \textit{persistent limitations in reporting on ecosystem and safety-related features of agentic systems}. 
These findings have implications for different stakeholders. For \textbf{policymakers}, the Index reveals that existing transparency expectations are largely unmet: most agents lack safety evaluations, disclosure mechanisms, and identity verification, suggesting that voluntary reporting is insufficient and structured requirements may be needed. For \textbf{developers}, the Index identifies concrete gaps---agent-specific system cards, sandboxing documentation, and web conduct policies---that should be addressed through stronger implementation and clearer disclosure. For \textbf{researchers}, the Index provides an empirical baseline for studying agent transparency, ecosystem concentration, and accountability fragmentation, and highlights the need for evaluation frameworks that target agentic behavior rather than model capabilities alone.

\subsection{Key Findings}

\textbf{The Index highlights inconsistent and selective reporting, particularly related to safety.} Developers rarely publish agent-specific evaluations. In the Index, only ChatGPT Agent, OpenAI Codex, Claude Code, and Gemini 2.5 Computer Use provide agent-specific system cards, though some Chinese agents have research papers focused on computer-use capabilities.
Meanwhile, only some companies report performance on \textit{capability} benchmarks (9/30). This transparency asymmetry suggests a weaker form of ``safety-washing'', where safety and ethics frameworks remain high-level and the empirical evidence required to rigorously assess risk is selectively disclosed \citep{biettiEthicsWashingEthics2020, floridiTranslatingPrinciplesPractices2019, renSafetywashingAISafety2024}.
This is potentially concerning because safety-critical behaviors emerge from planning, tools, memory, and policies rather than model capabilities alone. Agent builders frequently delegate some safety responsibilities to users rather than documenting built-in guardrails.

\textbf{Ecosystem-wide reliance on a few foundation models has implications for concentrated platform power.} Almost all systems in the Index rely on GPT-, Claude-, or Gemini-family models. Only foundation model developers based in the US and China operate their own proprietary models. This shared dependency creates potential single points of failure through pricing changes, service outages, and safety regressions \citep{kapoorPositionBuildAgent2025}. At the same time, the model-agnostic design of enterprise platforms may reduce lock-in risks, though this differs by market segment. This concentration of foundation models also potentially simplifies evaluation, as evaluators can focus resources on understanding the risks and capabilities of only a handful of models.

\begin{figure}
    \centering
    \includegraphics[width=1\linewidth]{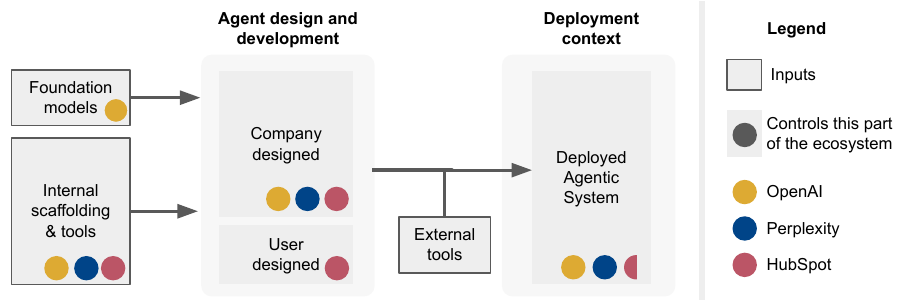}
    \Description{Block diagram of control across the agent ecosystem. Four boxes run left to right: foundation model, agent design and development, deployment context, and a small external tools box. Colored circles show which parts are controlled by OpenAI, Perplexity, and HubSpot. Control is split across stages, and no company controls the entire pipeline including external tools.}
    \caption{Control of parts of the AI agent ecosystem is fragmented, making reliable agent evaluation difficult. Individual developers often only control a subset of inputs (models, tools) and processes (agent design, deployment). For example, \ulchat{OpenAI’s ChatGPT Agent} controls the model and scaffolding and has visibility into the deployment context. \ulent{HubSpot’s Breeze Agent} controls orchestration and some inputs but may delegate agent design and configuration to users and may have only partial visibility and control of downstream deployment due to licensing constraints. \ulbrowser{Perplexity’s Comet} browser has direct access to the deployment environment. All companies have limited control over interactions with external tools.}
    \label{fig:ecosystem}
\end{figure}

\textbf{Evaluation of actual agentic risks is difficult across layers of the agent ecosystem.} Most agents rely on foundation models from frontier AI companies with scaffolding and orchestration layers built on top. \Cref{fig:ecosystem} shows how this architecture creates a chain of dependencies from model providers to orchestration platforms to agent builders to end-use deployments. Agentic evaluations inherently depend on the specific downstream context, including which tools are available and what level of autonomy the system has \citep{salaudeenMeasurementMeaningValiditycentered2025}. The lack of deployment-specific information makes it difficult to construct valid agent evaluations at the model level. Instead, evaluations should target tools in use, not just conversational safety. Regulators and buyers may risk false assurance from model-only documentation as the distributed architecture creates accountability diffusion where no single entity bears clear responsibility \citep{cooperAccountabilityAlgorithmicSociety2022}. This suggests the need for more information and risk sharing across the ecosystem, especially as capabilities are advancing faster than risk management practices \citep{ukaisecurityinstituteFrontierAITrends2025}.

\textbf{Agents' role on the web remains unsettled and potentially in tension with existing web scraping norms.} Browser-based agents often ignore robots.txt to function and appear to be designed to bypass anti-bot systems. Companies justify this by arguing that agents act directly on behalf of users and thus should not be subject to scraping restrictions \citep{perplexityteamAgentsBotsMaking2025}. 
This trend of agents bypassing robots.txt shifts control away from content hosts, suggesting that established web protocols may no longer be sufficient to mediate consent in an agentic ecosystem \citep{chan2025infrastructureaiagents, ziffdavis2025openai}.
Alternative governance mechanisms such as allowlisting frameworks and cryptographic authentication offer potential paths forward \citep{openai_agent_allowlisting, marroPermissionManifestsWeb2026}.
This tension is now being actively litigated, with platforms suing AI companies for bypassing technical controls (e.g., \citep{reddit2025anthropic, nyt2023openai,dowjones2024perplexity}) and challenging the legitimacy of ``agentic'' interactions (e.g., \citep{amazon2025perplexity}).
ChatGPT Agent is the only system in the Index to use cryptographic signing of requests. The absence of such signing makes it significantly harder to verify agent identity or to prove what an agent actually did, which may become increasingly important as more actions are delegated to agents. Similarly, the delegation of AI disclosure responsibilities from developers to operators raises questions about whether end users will be informed they are interacting with an AI, particularly when operators lack awareness of or incentive to comply with disclosure expectations.

\subsection{Limitations and Outlook}\label{sec:limitations}

\textbf{The Index's scope and methodology face limitations.} The AI agent ecosystem remains fundamentally difficult to document. Information is inconsistently available and reported. This difficulty has persisted since the inaugural AI Agent Index \citep{casper2025ai} and will likely continue absent structured reporting requirements or large-scale coordinated industry efforts.
Our inclusion criteria favor the most significant agents, which may affect generalizability. Smaller or emerging agents may exhibit different transparency patterns or introduce novel risks not captured here. Public interest metrics favor consumer products over enterprise deployments. Domain-specific agents are excluded. We prioritize high-quality human annotations over broad coverage; using language models with search capabilities may enable broader coverage \citep{longpreLargescaleAuditDataset2024, liangSystematicAnalysis321112024}.
The Index relies exclusively on publicly available information, which may miss internal evaluations or risk management practices. Further, the Index relies on English and Mandarin documentation and may miss information available in other languages, which may lead to understating transparency or safety practices for agents with documentation primarily in other languages. Finally, the Index may omit qualifying systems or contain inaccuracies despite vetting efforts and presents a snapshot as of December 31, 2025. We are committed to fixing errors on an ongoing basis, which can be shared via \url{https://aiagentindex.mit.edu/feedback}.

\textbf{Finally, the 2025 AI Agent Index raises open questions about the current AI agent ecosystem}, which we hope can be addressed in future work. 
Our analysis focuses on agentic systems that are publicly available, deployable with minimal configuration, and general-purpose. However, other systems, particularly ones deployed behind closed doors within frontier AI companies, remain much more opaque \citep{stix2025ai}. 
While some of the indexed agents will evolve or be superseded, the structural patterns identified here (foundation model concentration, accountability fragmentation, and capability-safety transparency gaps) are unlikely to resolve on their own. 
Likewise, the governance challenges documented (ecosystem fragmentation, web conduct tensions, absence of agent-specific evaluations) will gain importance as capabilities increase \citep{ukaisecurityinstituteFrontierAITrends2025}.
The Index provides a baseline against which future transparency improvements or regressions can be measured.
Future work could extend coverage to internal and domain-specific agents, more critically audit and compare the technical practices, reporting, and risk management of prominent agent developers, and track how these patterns evolve as governance frameworks mature.

\clearpage
\section*{Generative AI Usage Statement}

LLM outputs have not substantially contributed to the content or writing of this paper. Their usage was limited to the following tasks:

\begin{enumerate}
    \item \textbf{Candidate agent discovery:} ChatGPT 5.2 with deep research, Claude Sonnet 4.5 with research mode, and Gemini 2.5 with research mode were used to surface an initial list of candidate AI agents for potential inclusion (detailed prompts in \Cref{app:prompts_deep_research}). Human experts made final inclusion decisions based on criteria in \Cref{sec:inclusion_criteria}.

    \item \textbf{Annotation verification:} OpenAI GPT-5.2 with web search was used to cross-check human annotations for factual accuracy by searching for primary sources (see \Cref{app:verify}). All LLM-generated verification suggestions were manually reviewed and sources verified by human annotators.

    \item \textbf{Literature search assistance:} LLMs (Claude Sonnet 4.5, ChatGPT 5.2) were used to surface potentially relevant related work. All citations were independently verified and evaluated for relevance by authors.

    \item \textbf{Code generation for visualizations:} Multiple authors used Claude Code, Claude Opus 4.5, and Google Antigravity to generate Python code for data visualization and figure creation. All generated code was reviewed and verified by authors.

    \item \textbf{Copy-editing only:} Authors wrote all content and used Claude Sonnet 4.5, Claude Opus 4.5, and ChatGPT 5.2 only to correct grammar, fix typos, and improve clarity of existing sentences.
\end{enumerate}

\section*{Ethical Considerations Statement}

This work documents publicly available information about deployed AI agents. We considered the following ethical concerns and mitigation approaches:

\textbf{Privacy and data handling.} We collected only publicly available information. No sensitive user data, proprietary information, or private communications were analyzed. All web sources were archived for verification. Company correspondence was treated as confidential.

\textbf{Responsible disclosure and developer engagement.} Before publication, developers were given four weeks to review and correct annotations, with ongoing correction mechanisms available post-publication. This reduces risks of factual errors while maintaining research independence. 

\textbf{Potential for safety-washing or misrepresentation.} Our documentation of transparency gaps and safety practices could be selectively cited to misrepresent agent capabilities or inappropriately legitimize insufficient safety measures. We mitigate this by distinguishing ``None found'' (absence of public information) from ``None'' (confirmed absence), providing full context in annotations, and making the complete Index publicly available to enable verification.

\textbf{Resource and coverage biases.} Our significance criteria (search volume, market capitalization, developer prominence) favor well-funded companies and established products, potentially disadvantaging emerging developers and regional innovations. We mitigated this through consultation with Chinese ecosystem experts, multilingual search terms, and cross-referencing multiple agent databases. 

\textbf{Potential harm from publicizing security gaps.} Documenting safety and transparency limitations could guide malicious actors toward vulnerable systems. However, we report only publicly available information about major commercial systems without conducting novel security research or disclosing previously unknown vulnerabilities. Known incidents documented in the Index have already been publicly reported by security researchers or affected parties. We believe the benefits of transparency for governance and informed deployment decisions outweigh potential risks from documenting already-public information.

\begin{acks}
This research was supported by the MATS Research program, which provided funding for L.S. and M.Y. through research stipends. We also thank MATS and our research manager Keivan Navaie for their organizational assistance and research support.

We are grateful to Alan Chan, Kevin Klyman, Lily Stelling, Robert Adragna, and Anna Schuh for their valuable feedback on earlier drafts of this paper. We also thank participants of the Partnership on AI workshop on monitoring and the UK AI Forum workshop for helpful discussions that shaped this work. We thank Xudong Pan for help in verifying our research on Chinese agents and pointing us to additional resources.
\end{acks}

\section*{Contribution Statement}
L.S. led the project, developed the methodology, led data collection and analysis, coordinated agent annotations, created visualizations, and co-wrote the paper.

M.Y., L.B., K.F., K.W., A.P.O., and Y.D. contributed to agent annotations and data curation. Y.D. led annotations for Chinese agents.
S.C. and N.K. supervised the project, contributed to conceptualization and methodology, and co-wrote the paper.
All authors reviewed the final paper.

\bibliographystyle{ACM-Reference-Format}
\bibliography{aiaiv1,literature_new}

\clearpage
\appendix

\section{The 2025 AI Agent Index} \label{app:index}
The 2025 AI Agent Index is available at: \url{https://aiagentindex.mit.edu}

The full annotations for all fields are available in JSON and CSV format on Zenodo at: \url{https://doi.org/10.5281/zenodo.18701930}

\begin{figure}[H]
    \centering
    \includegraphics[width=1\linewidth]{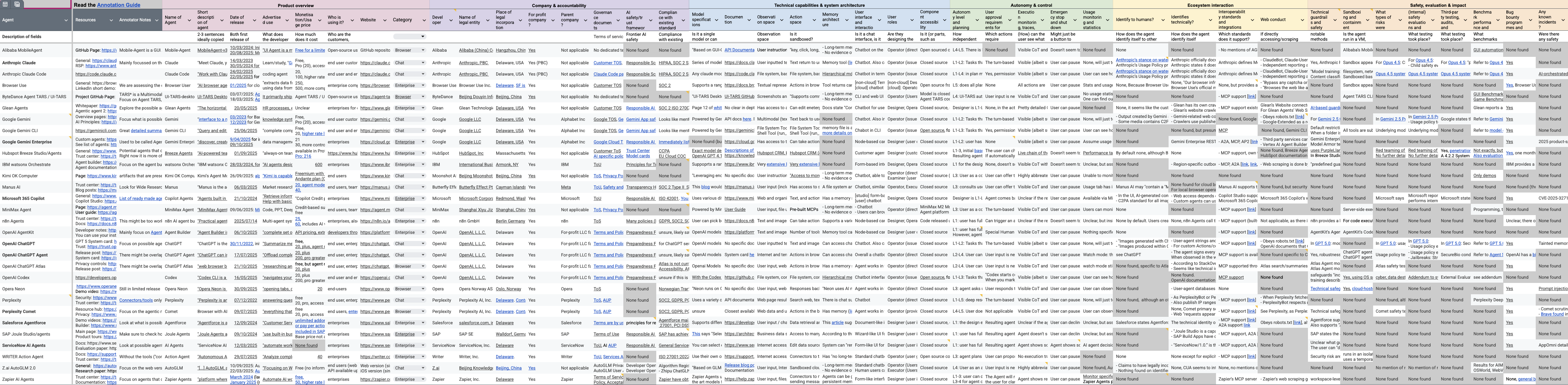}
    \Description{Screenshot of the AI Agent Index spreadsheet. Rows list agent products and columns span grouped fields for product overview, company and accountability, technical capabilities, autonomy and control, ecosystem interaction, and safety, evaluation and impact. The sheet is very wide, with dense text entries, links, and missing-value markers in many cells.}
    \caption{The 2025 AI Agent Index with detailed annotations across 6 categories (45 columns) for 30 agentic AI products.}
    \label{fig:agent_index}
\end{figure}

\subsection{Further Analysis}
\begin{figure}[H]
    \centering
    \includegraphics[width=1\linewidth]{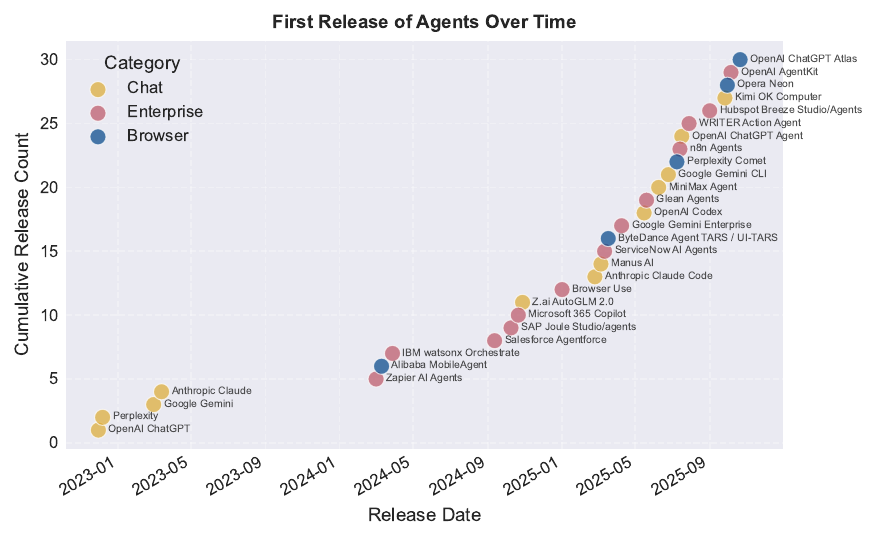}
    \Description{Scatterplot timeline of first agent releases. The horizontal axis is release date from early 2023 to late 2025 and the vertical axis is cumulative release count. Colored points by category step upward over time, with only a few releases in 2023 and a dense cluster of releases through 2025. Most points in the upper half of the chart occur in 2025.}
    \caption{First release of indexed agentic AI products over time and by agent category (\ulchat{chat}, \ulent{enterprise}, and  \ulbrowser{browser}).}
    \label{fig:agent_release_timeline}
\end{figure}

\begin{figure}[H]
    \centering
    \includegraphics[width=0.8\textwidth]{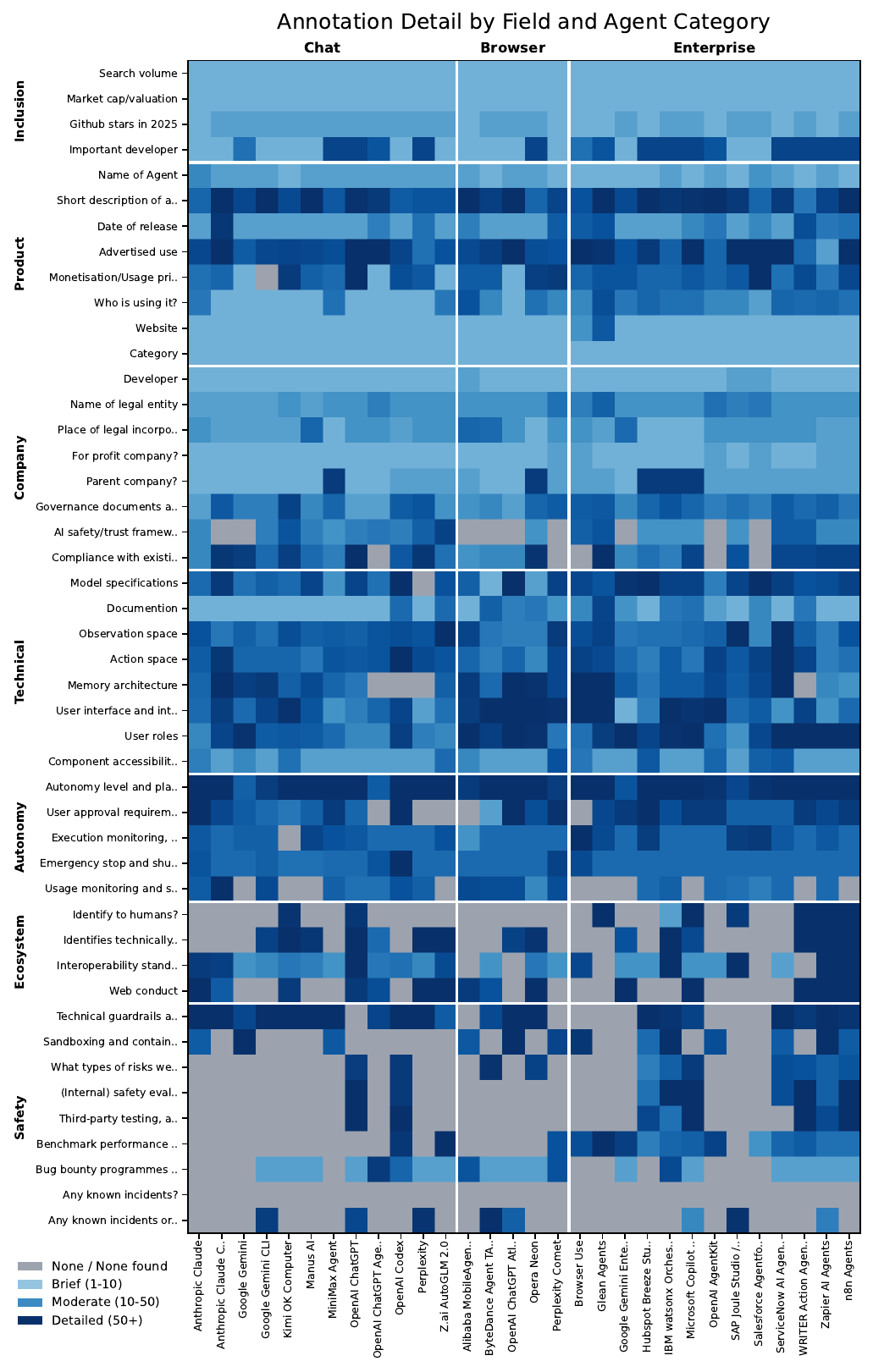}
    \Description{Full-page heatmap of annotation detail by field and agent. Rows are agents and columns are fields grouped by category. Gray marks none or none found and darker blue marks more detailed entries. Product, company, and technical sections are mostly blue, while ecosystem and safety include many gray cells across the chart.}
    \caption{For 227 out of 1,350 fields, we were unable to find any information (gray). This is most common in the ``Ecosystem Interaction'' and ``Safety, Evaluation, and Impact'' categories. Non-empty information fields are 14 words long on average.}
    \label{app:transparency_gaps_full}
\end{figure}

\begin{figure}[H]
    \centering
    \includegraphics[width=0.8\linewidth]{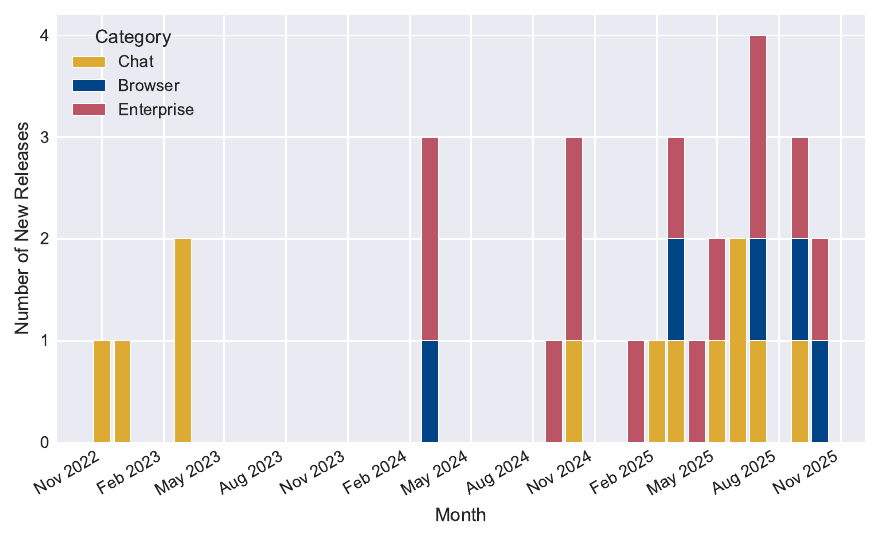}
    \Description{Stacked bar chart of new releases by month. The horizontal axis runs from late 2022 to late 2025 and the vertical axis shows the number of new releases. There are few bars before 2024, then repeated monthly releases through 2025. The tallest bar is around August 2025 with four releases, and most late bars combine chat, browser, and enterprise categories.}
    \caption{Number of new AI agentic product releases by month.}
    \label{fig:agent_releases_per_month}
\end{figure}

\begin{figure}[H]
    \centering
    \includegraphics[width=0.8\linewidth]{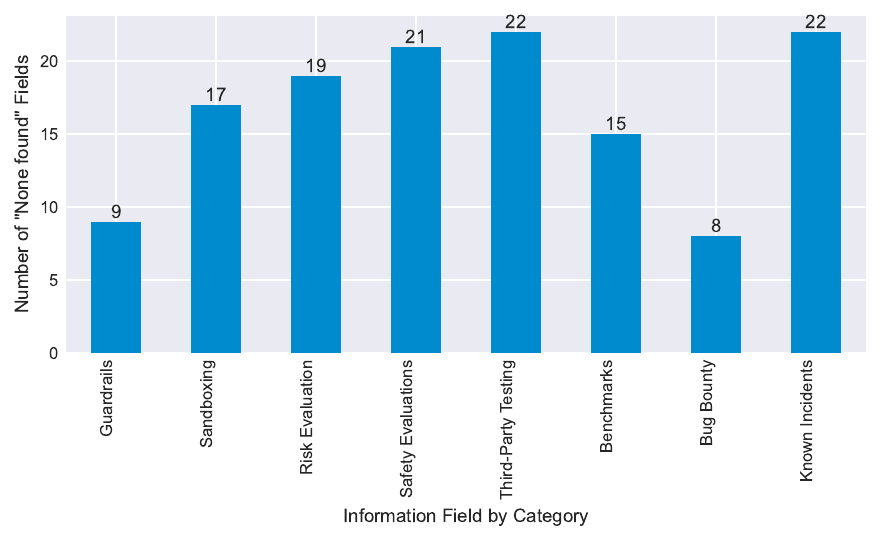}
    \Description{Bar chart of counts of none-found fields by safety information category. Third-party testing and known incidents are highest at 22 each, followed by safety evaluations at 21 and risk evaluation at 19. Bug bounty is the lowest at 8.}
    \caption{Number of information fields with ``None found'' by field category.}
    \label{fig:none_found_by_category}
\end{figure}

\begin{figure}[H]
    \centering
    \includegraphics[width=1\linewidth]{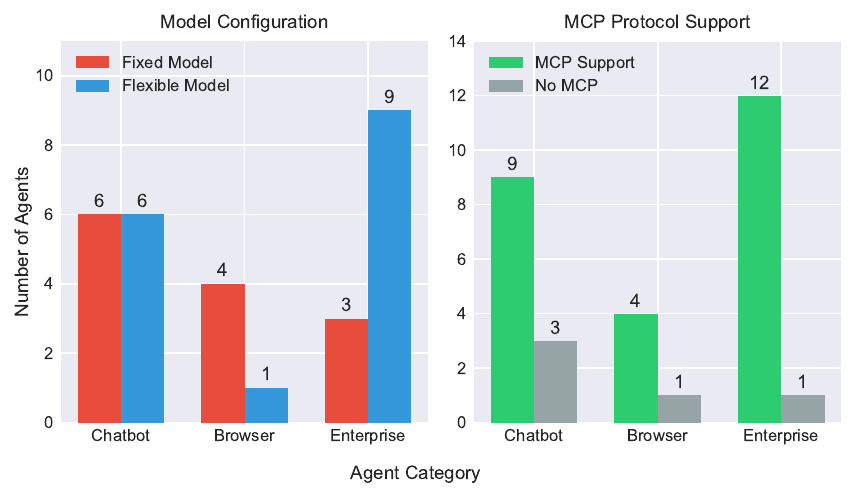}
    \Description{Two grouped bar charts comparing technical features by category. The left chart shows fixed versus flexible model configuration: chat is split evenly, browser is mostly fixed, and enterprise is mostly flexible. The right chart shows Model Context Protocol support: enterprise is highest at 12 of 13, chat has 9 of 12, and browser has 4 of 5.}
    \caption{Support for choosing models and MCP varies by category of the agent. Enterprise agents are more likely to support model selection (9/13) and MCP (12/13).}
    \label{fig:technical_by_category}
\end{figure}

\subsection{Sample Entry of the Index: Claude Code} \label{app:sample_entry}

This section provides a sample entry of the Index for Anthropic's Claude Code, demonstrating how each field is documented.
We selected it based on its high degree of documentation (no ``None'' or ``None found'' annotations).
No authors have conflicts of interest related to Anthropic or Claude Code, and this example selection was made without correspondence with Anthropic. 
Including Claude Code as an example here is not an endorsement of the system or developer.

\subsubsection*{Inclusion Criteria}
\mbox{}

\textbf{Search volume:} 348,625 peak monthly hits.

\textbf{Valuation:} USD 183 billion.

\textbf{GitHub stars in 2025:} \href{https://github.com/anthropics/claude-code}{50,700}.

\textbf{Important developer:} Yes. Included in Foundation Model Transparency Index. Member of Frontier Model Forum. Signatory of Frontier AI Safety Commitments.

\subsubsection*{Product Overview}
\mbox{}

\textbf{Name of Agent:} Claude Code.

\textbf{Short description of agent:} \href{https://claude.com/product/claude-code}{``Work with Claude directly in your terminal. Claude explores your codebase context, answers questions, and make changes''}.

\textbf{Date of release:} 24/02/2025 (initial release), 22/05/2025 (\href{https://claudelog.com/faqs/claude-code-release-notes/\#v100}{general release}).

\textbf{Advertised use:} Coding agent.

\textbf{Monetization/Usage price:} \$20/month (Pro), \$100/month (Team with higher rate limits), \$200/month (Enterprise).

\textbf{Who is using it:} End user and enterprise customers for coding and prototyping.

\textbf{Website:} \url{https://claude.com/product/claude-code}.

\textbf{Category:} Chat.

\subsubsection*{Company \& Accountability}
\mbox{}

\textbf{Developer:} Anthropic.

\textbf{Name of legal entity:} \href{https://www.anthropic.com/legal/commercial-terms}{Anthropic, PBC}.

\textbf{Place of legal incorporation:} Delaware, USA.

\textbf{For profit company:} Yes (PBC).

\textbf{Parent company:} Not applicable.

\textbf{Governance documents analysis:} \href{https://code.claude.com/docs/en/legal-and-compliance}{Claude Code page}, \href{https://www.anthropic.com/legal/consumer-terms}{Customer TOS}, \href{https://www.anthropic.com/legal/aup}{Usage Policy}.

\textbf{AI safety/trust framework:} \href{https://www-cdn.anthropic.com/872c653b2d0501d6ab44cf87f43e1dc4853e4d37.pdf}{Responsible Scaling Policy}.

\textbf{Compliance with existing standards:} \href{https://trust.anthropic.com/}{HIPAA, SOC 2 Type I and II, ISO 27001:2022, ISO/IEC 42001:2023, FedRAMP High, UK Cyber Essentials}.

\subsubsection*{Technical Capabilities \& System Architecture}
\mbox{}

\textbf{Model specifications:} Any Claude model. Default depends on subscription tier. \href{https://code.claude.com/docs/en/model-config}{User can choose model}.

\textbf{Documentation:} \url{https://code.claude.com/docs}.

\textbf{Observation space:} File system, bash commands, MCP.

\textbf{Action space:} File system, bash commands, MCP.

\textbf{Memory architecture:} \href{https://code.claude.com/docs/en/memory}{Hierarchical markdown memory}.

\textbf{User interface and interaction design:} Chatbot in terminal.

\textbf{User roles:} Operator (issues queries, which the agent responds to); Executor (user may take actions/make decisions based on outputs); Examiner (user can use thumbs up/down buttons to give feedback).

\textbf{Component accessibility:} Closed source.

\subsubsection*{Autonomy \& Control}
\mbox{}

\textbf{Autonomy level and planning depth:} L1-L4: in plan mode it is most like a simple chatbot, but with auto-approve mode on, Claude Code can plan actions and take multiple steps (using different tools) without user approval. It will ask for clarification as needed.

\textbf{User approval requirements:} Yes, permission for running bash commands, editing files, or reading files outside of its initial directory source.

\textbf{Execution monitoring, traces, and transparency:} Visible (albeit summarized) chain-of-thought with a list of to-dos being worked on.

\textbf{Emergency stop and shutdown mechanisms:} User can pause/stop the agent at any time.

\textbf{Usage monitoring and statistics:} User can see how much context is used.

\subsubsection*{Ecosystem Interaction}
\mbox{}

\textbf{Identify to humans:} \href{https://www.anthropic.com/transparency/voluntary-commitments\#:~:text=output\%2E-,While,area}{Anthropic's stance on watermarking}: ``While watermarking is most commonly applied to image outputs, which we do not currently provide, we continue to work across industry and academia to explore and stay abreast of technological developments in this area.'' Anthropic's Usage Policy prohibits using Claude to impersonate a human (i.e., to convince someone they're communicating with a natural person when they are not), implying Claude deployments must not hide AI identity in human interactions \href{https://terms.law/2025/04/09/navigating-ai-platform-policies-who-owns-ai-generated-content/}{link}.

\textbf{Identifies technically:} Anthropic officially documents that Claude-related web activity is identifiable via specific User-Agent tokens: ClaudeBot, Claude-User, and Claude-SearchBot \href{https://support.anthropic.com/en/articles/8896518-does-anthropic-crawl-data-from-the-web-and-how-can-site-owners-block-the-crawler}{link}. Anthropic states it does not currently publish fixed IP ranges for these bots/agents (they use service-provider public IPs), so IP-range identification is not available as an official signature mechanism \href{https://support.anthropic.com/en/articles/8896518-does-anthropic-crawl-data-from-the-web-and-how-can-site-owners-block-the-crawler}{link}.

\textbf{Interoperability standards and integrations:} Anthropic defines Model Context Protocol (MCP) as an open standard that ``standardizes how applications provide context to LLMs,'' likened to a ``USB-C port for AI applications'' \citep{WhatModelContext}. Also works with open source plugins and skills.

\textbf{Web conduct:} ClaudeBot, Claude-User, and Claude-SearchBot ``respect 'do not crawl' signals by honoring industry standard directives in robots.txt'' and ``respect anti-circumvention technologies,'' stating they do not attempt to bypass CAPTCHAs \href{https://support.claude.com/en/articles/8896518-does-anthropic-crawl-data-from-the-web-and-how-can-site-owners-block-the-crawler}{link}. Independent reporting and site-operator accounts, however, have documented periods of very heavy crawling and, at least in some cases, behavior that appeared to ignore site preferences until new robots.txt rules propagated \href{https://www.businessinsider.com/openai-anthropic-ai-ignore-rule-scraping-web-contect-robotstxt}{link} \href{https://www.theverge.com/2024/7/25/24205943/anthropic-ai-web-crawler-claudebot-ifixit-scraping-training-data}{link}.

\subsubsection*{Safety, Evaluation \& Impact}
\mbox{}

\textbf{Technical guardrails and safety measures:} Model training using reinforcement learning to recognize and refuse malicious instructions. Content classifiers scan untrusted content and flag potential injections. Granular permissions to control what Claude can access and do. Site blocklists preventing access to certain high-risk websites. Action confirmations for high-risk actions such as purchasing \href{https://support.claude.com/en/articles/12902428-using-claude-in-chrome-safely}{link}. Claude Code uses a permission-based architecture and is read-only by default; higher-impact actions (e.g., editing files, executing commands) require explicit user approval \href{https://docs.claude.com/en/docs/claude-code/security}{link}.

\textbf{Sandboxing and containment approaches:} Filesystem, network, OS-level enforcement. Sandboxed bash tool: Sandbox bash commands with filesystem and network isolation, reducing permission prompts while maintaining security. Enable with \texttt{/sandbox} to define boundaries where Claude Code can work autonomously. Write access restriction: Claude Code can only write to the folder where it was started and its subfolders \href{https://code.claude.com/docs/en/sandboxing}{source}.

\textbf{What types of risks were evaluated:} \href{https://assets.anthropic.com/m/64823ba7485345a7/Claude-Opus-4-5-System-Card.pdf}{Opus 4.5 system card} contains section on agentic misuse.

\textbf{(Internal) safety evaluations and results:} \href{https://assets.anthropic.com/m/64823ba7485345a7/Claude-Opus-4-5-System-Card.pdf}{Opus 4.5 system card} contains section on agentic misuse.

\textbf{Third-party testing, audits, and red-teaming:} \href{https://assets.anthropic.com/m/64823ba7485345a7/Claude-Opus-4-5-System-Card.pdf}{Opus 4.5 system card}, section 5.2.1 Gray Swan Agent Red Teaming benchmark for tool use.

\textbf{Benchmark performance and demonstrated capabilities:} Refer to \href{https://assets.anthropic.com/m/64823ba7485345a7/Claude-Opus-4-5-System-Card.pdf}{Opus 4.5 System Card}.

\textbf{Bug bounty programs and vulnerability disclosure:} \href{https://hackerone.com/anthropic-vdp/reports/new?type=team&report_type=vulnerability}{Yes}.

\textbf{Any known incidents or reported vulnerabilities:} \href{https://www.anthropic.com/news/disrupting-AI-espionage}{AI-orchestrated cyber espionage campaign}.

\section{Annotation Methodology}

\subsection{List of Agentic AI Products Considered} \label{app:full_agent_list}
Below is the complete list of agent products considered, with those included in the final 2025 AI Agent Index in bold. Some enterprise agent products also include ready-made agents (e.g. Microsoft 365 Copilot). Our analysis does not focus on these. Instead, we focus on agentic AI products that allow for general customization (e.g. Microsoft Copilot Studio) and the agents that could be created through these.
\begin{multicols}{2}
\begin{itemize}
    \item 01.AI WorldWise Enterprise LLM Platform
    \item Aider
    \item AI21 Maestro
    \item \textbf{Alibaba MobileAgent}
    \item All Hands OpenHands
    \item Aleph Alpha Pharia
    \item Amazon Bedrock Agents
    \item Amazon Nova Act
    \item Amazon Q Business
    \item Anysphere Cursor
    \item \textbf{Anthropic Claude}
    \item \textbf{Anthropic Claude Code}
    \item Automation Anywhere AI Agents
    \item Baidu Agent Platforms
    \item Beam AI Beam
    \item \textbf{Browser Use}
    \item \textbf{ByteDance Agent TARS}
    \item ByteDance Coze
    \item ByteDance TRAE
    \item Cline
    \item Cloudflare Agents
    \item Cognition Devin
    \item Cognition Windsurf
    \item Cohere North
    \item Continue (Continue.dev)
    \item Counsel AI Corporation Harvey Agents
    \item CrewAI Crew
    \item Databricks Agent Bricks
    \item Determinist Ltd AutoGPT
    \item Dust
    \item Flowise
    \item Genspark Super Agent
    \item GitHub Copilot Agent
    \item \textbf{Glean Agents}
    \item \textbf{Google Agentspace}
    \item \textbf{Google Gemini}
    \item \textbf{Google Gemini Code Assist}
    \item Google DeepMind Project Astra
    \item Google Jules
    \item Google Project Mariner
    \item Gumloop AI Workflows
    \item \textbf{HubSpot Breeze Agents}
    \item HuggingFace Computer Agent
    \item \textbf{IBM watsonx Orchestrate}
    \item Kiln AI Kiln Agents
    \item Kiro
    \item Lindy
    \item Lovable
    \item \textbf{Manus AI Manus}
    \item \textbf{Microsoft Copilot Studio}
    \item Microsoft Magentic UI
    \item MindStudio AI Agents
    \item \textbf{MiniMax}
    \item Model scope MS Agent
    \item \textbf{Moonshot AI Kimi OK Computer}
    \item Motion AI Employee
    \item Moveworks Agent studio
    \item \textbf{n8n}
    \item NAVER Cue Search
    \item Notion Agent
    \item \textbf{OpenAI AgentKit}
    \item OpenAI Agent SDK
    \item \textbf{OpenAI ChatGPT}
    \item \textbf{OpenAI ChatGPT Agent}
    \item \textbf{OpenAI ChatGPT Atlas}
    \item \textbf{OpenAI Codex}
    \item OpenManus
    \item \textbf{Opera Neon}
    \item Oracle AI Agent Platform
    \item OutSystems Agent Workbench
    \item Palantir AIP
    \item Pega Blueprint
    \item \textbf{Perplexity}
    \item \textbf{Perplexity Comet}
    \item Relevance AI agents
    \item Replit Agent
    \item Sakana AI Scientist
    \item \textbf{Salesforce Agentforce}
    \item \textbf{SAP Joule studio}
    \item \textbf{ServiceNow AI Agents}
    \item Sierra Agent
    \item Skyvern
    \item Sourcegraph Cody
    \item Spell
    \item StackAI
    \item StackBlitz Bolt
    \item SWE-agent
    \item Tencent AppAgent
    \item Tencent Youtu-Agent
    \item The San Francisco AI Factory Inc Factory
    \item UiPath Autopilot
    \item Workday AI Agents
    \item \textbf{WRITER Action Agent}
    \item \textbf{Z.ai AutoGLM}
    \item \textbf{Zapier AI Agents}
    \item LangChain LangSmith Agent Builder
\end{itemize}
\end{multicols}

\clearpage
\subsection{Annotation Fields} \label{app:annotation_fields}

\textbf{1. Inclusion Criteria}
\begin{itemize}
  \item \textbf{Search volume:} Monthly search estimates for top 5 keywords using Ahrefs to measure public interest over 2025.
  \item \textbf{GitHub stars in 2025:} Stars for GitHub repositories about the agent product itself (not related ``cookbooks'' or similar), where applicable.
  \item \textbf{Market capitalization/valuation:} Developer market cap as a December 2025 average (public companies) or valuation as of December 2025 (private companies). Data from Yahoo Finance, Crunchbase, Epoch AI, and general news sources.
  \item \textbf{Important developer:} Membership in 2024 Foundation Model Transparency Index \citep{bommasani20242024}, Frontier Model Forum \citep{frontiermodelforumMembership}, or signatory of Frontier AI Safety Commitments \citep{FrontierAISafety2025} or Artificial Intelligence Safety Commitments \citep{chinaacademyofinformationandcommunicationstechnologyFirst17Companies2024}.
\end{itemize}

\textbf{2. Product Overview}
\begin{itemize}
  \item \textbf{Name of Agent}
  \item \textbf{Short description:} 2-3 sentence description copied directly from developer (main marketing headline).
  \item \textbf{Date of release:} First release and latest update (month-level granularity).
  \item \textbf{Advertised use:} Developer-stated capabilities and intended use cases. Category of use (finance, web development) or specific examples (CRM to prioritize leads, summarize sales from different sources).
  \item \textbf{Monetization/Usage price:} Cost per month per user/seat in USD. Subscription tiers. Access method if not directly monetized (e.g., part of existing Microsoft/Google subscription). Additional costs (API model calls, storage).
  \item \textbf{Who is using it:} Customer types (end users, enterprises by size/industry, through API, governments).
  \item \textbf{Website:} Product landing page.
  \item \textbf{Category}: ``Chatbot with tools,'' ``Browser-based,'' or ``Enterprise''. See \Cref{sec:whats_included} for details.
\end{itemize}

\textbf{3. Company \& Accountability}
\begin{itemize}
  \item \textbf{Developer:} Developer name.
  \item \textbf{Name of legal entity:} Legal entity name, location of headquarters, legal domicile, data residency (including state if US).
  \item \textbf{Place of legal incorporation:} Headquarters location, legal domicile.
  \item \textbf{For profit company:} Corporate structure (for-profit, public benefit corporation, other structures).
  \item \textbf{Parent company:} Parent company ownership if applicable.
  \item \textbf{Governance documents analysis:} Terms of Service, privacy policy, acceptable use policy/usage restrictions, AI-specific policy.
  \item \textbf{AI safety/trust framework:} Responsible Scaling Policy (RSP), Frontier AI Safety Framework, company-wide safety documentation.
  \item \textbf{Compliance with existing standards:} Claimed compliance with ISO/IEC, NIST AI RMF, EU AI Act categories, SOC, relevant laws (e.g., GDPR).
\end{itemize}

\textbf{4. Technical Capabilities \& System Architecture}
\begin{itemize}
  \item \textbf{Model specifications:} Whether single model or user-selectable models. Available models by developer. Foundation model name, checkpoint if available. Whether reasoning model.
  \item \textbf{Documentation:} Links to technical documentation.
  \item \textbf{Observation space:} Input information sources. Internet access. Model Context Protocol (MCP) support.
  \item \textbf{Action space:} Sandboxing status. Whether an agent can directly affect the real world without human approval or be configured to. Available tools with write access.
  \item \textbf{Memory architecture:} Types employed (short-term, long-term, consensus, episodic) \citep{googlecloudWhatAreAI2025}.
  \item \textbf{User interface and interaction design:} Interface type (chat, browser integration, other application). Whether anthropomorphism encouraged. Warnings against misperception. Human--agent teaming practices (overreliance mitigation, error communication).
  \item \textbf{User roles:} Designer, Operator, Executor, Examiner capabilities (designing agent, running and providing inputs, interacting with output and making decisions, evaluating through feedback), as per \citet{tomsettInterpretableWhomRolebased2018}.
  \item \textbf{Component accessibility:} Open-source status and license. Availability of weights, data, code, scaffolding.
\end{itemize}

\textbf{5. Autonomy \& Control}
\begin{itemize}
  \item \textbf{Autonomy level and planning depth:} L1-L5 classification as per \citet{fengLevelsAutonomyAI2025}: L1 (user as operator, agent provides on-demand support), L2 (user as collaborator, agent works independently on own tasks), L3 (user as consultant, agent takes initiative over extended time horizons), L4 (user as approver, interaction only when agent encounters blockers), L5 (user as observer, no means for user involvement).
  \item \textbf{User approval requirements:} Which actions require explicit approval? Whether approval is given once or for all following actions. Whether approval can be revoked.
  \item \textbf{Execution monitoring, traces, and transparency:} How users can see agent actions. Whether in real time, as a recording afterwards, or in another format.
  \item \textbf{Emergency stop and shutdown mechanisms:} Stop/abort controls. Relevant for agents running continuously or based on triggers. Whether builder platforms allow creating shutdown mechanisms to delegate control back to the user.
  \item \textbf{Usage monitoring and statistics:} Activity tracking, usage patterns.
\end{itemize}

\textbf{6. Ecosystem Interaction}
\begin{itemize}
  \item \textbf{Identify to humans:} Whether agent identifies as AI when interacting with non-user humans. Provenance tracking for outputs (e.g., watermarking).
  \item \textbf{Identifies technically:} Digital signatures the agent uses. Published IP ranges, user agent strings, cryptographic signatures for web requests, request signing, unique identifiers.
  \item \textbf{Interoperability standards and integrations:} Supported standards and frameworks for agent communication. Examples include AGNTCY, Agent Connect Protocol (ACP), Model Context Protocol (MCP), Agent2Agent (A2A) protocol. Whether agent has own API. Whether agents use MCP to access tools or have MCP servers for others to use their services.
  \item \textbf{Web conduct:} Whether agent complies with robots.txt when directly accessing/scraping web. Crawling behavior. Anti-bot evasion techniques.
\end{itemize}

\textbf{7. Safety, Evaluation \& Impact}
\begin{itemize}
  \item \textbf{Technical guardrails and safety measures:} Notable methods used to protect against harmful actions. Built-in guardrails. For agent builders, types of guardrails that can be added and conditions under which they are active.
  \item \textbf{Sandboxing and containment approaches:} Whether agent runs in virtual machine (VM), locally, or other environment. Characteristics of VM and how it interacts with environment. For hosted products, specific details on containment practices beyond general infrastructure.
  \item \textbf{What types of risks were evaluated:} Scope of risk assessment.
  \item \textbf{(Internal) safety evaluations and results:} Testing scope and procedures. Whether evaluations are agent-specific or model-only. Results of evaluations.
  \item \textbf{Third-party testing, audits, and red-teaming:} External testing scope and organizations involved. Specific to agent (does not include general compliance audits).
  \item \textbf{Benchmark performance and demonstrated capabilities:} Benchmarks run and results (developer-reported only).
  \item \textbf{Bug bounty programs and vulnerability disclosure:} Links to programs if applicable. Disclosure policies.
  \item \textbf{Any known incidents or reported vulnerabilities:} Safety incidents in 2025.
\end{itemize}

\subsection{Changes in Annotation Fields} \label{app:changes_annotation_fields}
\begin{figure}[H]
    \centering
    \includegraphics[width=1\linewidth]{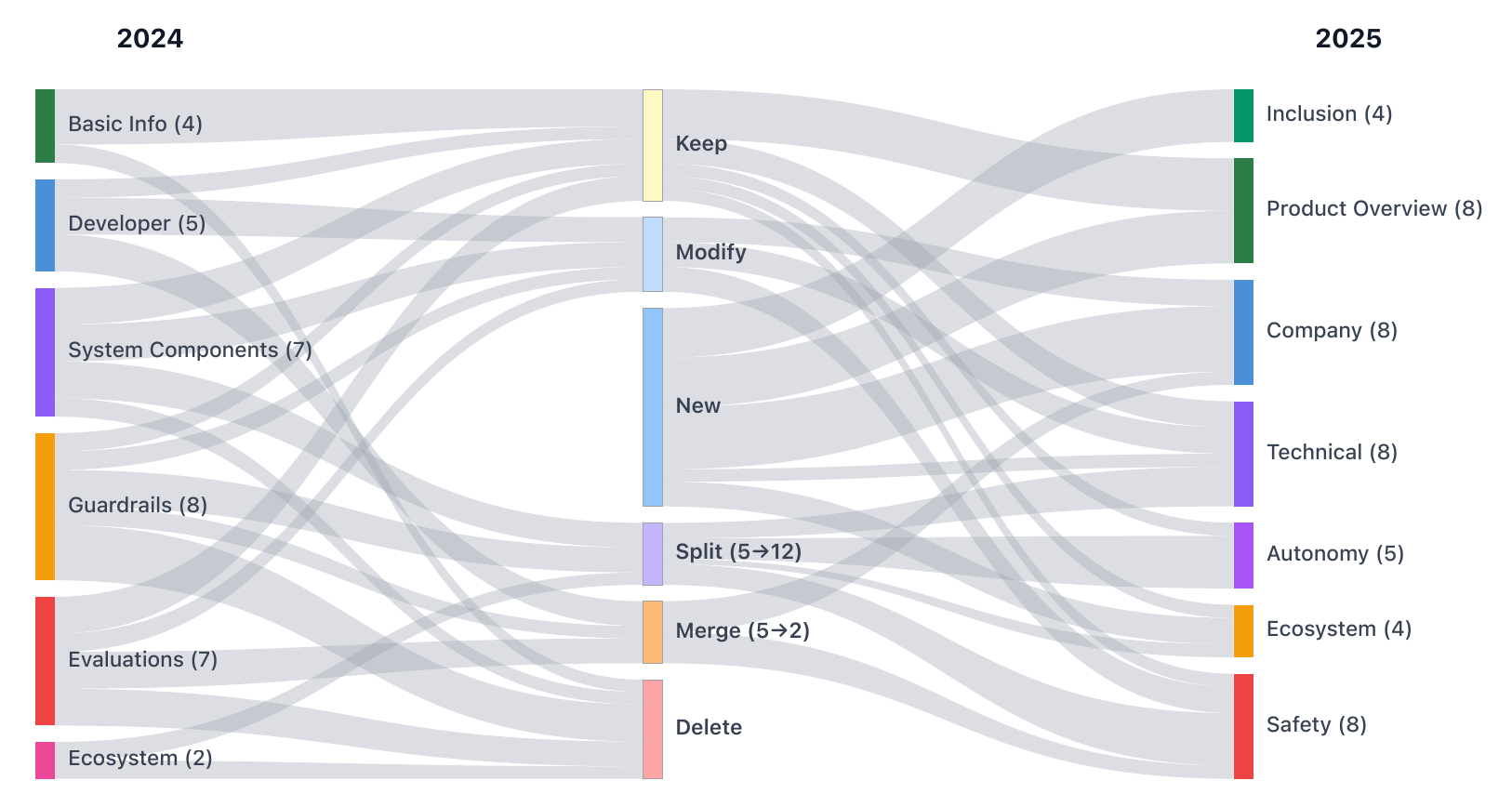}
    \Description{Sankey diagram of annotation field changes from 2024 to 2025. Category groups from 2024 appear on the left and 2025 groups on the right, connected through flows labeled keep, modify, and new in the center. Only a small share of fields flow through keep, while larger flows go through modified or newly added fields.}
    \caption{\textbf{Changes in annotation fields across categories since the 2024 Index.} We make significant changes across all sections compared to the 2024 Index \citep{casper2025ai}: 16 fields in the 2025 version are completely new, and only 9 fields are kept unaltered (\textit{Keep}). The remaining 16 fields were derived from fields in the 2024 version either by making significant modifications to the definition/notes (\textit{Modify}), splitting a field (\textit{Split}), or merging multiple fields into one (\textit{Merge}). 8 fields were deleted.}
    \label{fig:changes_annotation_fields}
\end{figure}

\subsection{Annotation Guide} \label{app:annotation_guide}

Annotators received detailed instructions emphasizing the following principles:

\textbf{Depth and detail.} With approximately 30 agents, conduct thorough analysis through demo videos, secondary sources, and testing. Focus on object-level findings rather than interpretations. Document defaults chosen in UI or architecture. Include tangential observations and reasoning for educated guesses.

\textbf{Agent-specific focus.} Annotate safety and transparency features of agents themselves, not underlying models. Evaluations of foundation models alone (e.g., GPT-4o) do not count as agent evaluations unless properly contextualized.

\textbf{Platforms vs. agents.} For platforms creating agents rather than standalone agents, annotate based on most capable agents that can be created. Note default options and platform capabilities/limitations.

\textbf{Source documentation.} Cite sources for every field with multiple links when applicable. Link directly to highlighted text. Include timestamps for video citations.

\textbf{Handling uncertainty.} Use standardized markers: ``None found'' when confident information doesn't exist, ``Not applicable'' when field doesn't apply to the agent, ``UNSURE'' when information might exist somewhere, ``TODO'' for later revisits. Liberally use comments for second opinions.

\textbf{Information sources.} Official documentation, company blog posts, help center documents, trust center materials (including penetration test reports), and developer conference demo videos. Annotators shared helpful sources for collective benefit.

\textbf{Scope of information.} Annotations based on publicly available information, including what is visible in the agent interface itself. No experiments conducted (e.g., testing how agents identify themselves to websites). For open-source agents, documentation and readme files used rather than code analysis.

\subsection{LLM Prompts for Finding Agents} \label{app:prompts_deep_research}
We used LLM-based research queries to surface an initial list of candidate agents. Below we provide the prompts used for each given model.

ChatGPT 5.2 with deep research:
\begin{itemize}
    \item \lstinline|what are the most significant coding agents that you can use to do general purpose things. e.g. the coding agents must support MCP and be capable of taking non coding actions as well.| 
    \item \lstinline|I currently have this list of agents. are any important agents missing? output just a list of these agents with names and developer and link.|
    \item \lstinline|what are the most significant AI agents currently available. they should be actually agentic, not just AI chatbots.|
    \item \lstinline|summarise the recent developments in AI agents, what major new advanced Ai agent have come to the market in 2024 and 2025, what impact have they had. Also consider what academic literature on AI agents has come out recently.|
\end{itemize}

Claude Sonnet 4.5 with research mode:
\begin{itemize}
    \item \lstinline|what are the most significant AI agents currently available. they should be actually agentic, not just AI chatbots.|, followed by \lstinline|I am looking for all available real world examples that can be used right away, open source or commercial, across all domains (although domain specific is fine too). with actually agentic I mean all of these capabilities.|
    \item \lstinline|summarise the recent developments in AI agents, what major new advanced Ai agent have come to the market in 2024 and 2025, what impact have they had. Also consider what academic literature on AI agents has come out recently.|, followed by \lstinline|give me a list of significant AI agent products (both open source and industry) available right now. focus on those used by many people or from notable companies or that are talked about a lot. include everything. But mention if they are not publically released yet. if they are specialised agents they need to be more significant, eg. even more users. mention the level of autonomy.|
\end{itemize}

Gemini 2.5 with research mode:
\begin{itemize}
    \item \lstinline|what are the most significant AI agents currently available. they should be actually agentic, not just AI chatbots.|
    \item \lstinline|summarise the recent developments in AI agents, what major new advanced Ai agent have come to the market in 2024 and 2025, what impact have they had. Also consider what academic literature on AI agents has come out recently.|
\end{itemize}

\subsection{LLM Usage to Verify Annotations}\label{app:verify}
\begin{figure}[h]
    \centering
    \includegraphics[width=1\linewidth]{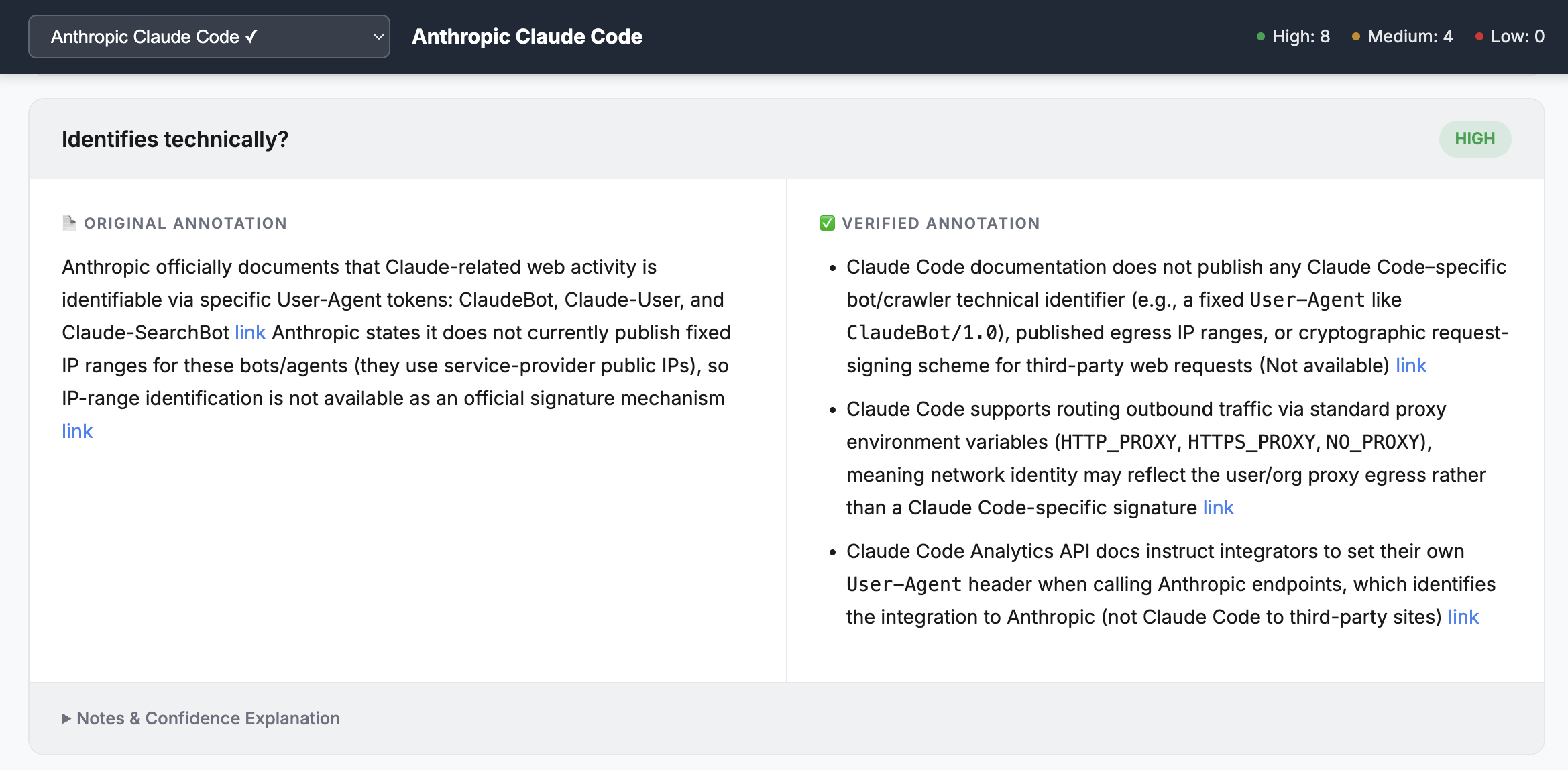}
    \Description{Screenshot of the annotation verification viewer. A dark header contains an agent selector and counts of high, medium, and low confidence results. The main panel shows the field ``Identifies technically?'' with original annotation text on the left and verified annotation text on the right, plus a green high-confidence badge.}
    \caption{Custom annotation viewer to compare the human annotations to the LLM-generated ones in the verification phase.}
    \label{fig:annotation_viewer}
\end{figure}

To enhance the reliability of our annotations, we developed an automated verification pipeline using OpenAI's GPT-5.2 model with web search capabilities. We provide all prompts used in the process below. The verification process follows a three-step approach:

\begin{enumerate}
    \item \textbf{Web research phase:} For each agent-field pair, the system prompts the model to search for primary sources (official documentation, press releases, developer blogs, and trust centers) related to the specific agent product. The prompt is constrained to match the annotation guide used by human annotators, see \Cref{app:annotation_guide}.
    
    \item \textbf{Structured verification phase:} The research findings are then processed by a second model call that outputs a structured JSON response containing: (a) a verified annotation with inline source citations, (b) a confidence rating (High/Medium/Low), (c) confidence justification, and (d) any discrepancies with the original annotation.

    \item \textbf{Manual verification phase:} An annotator compared the original annotations to the LLM-generated ones and, if necessary, amended the final annotation. Each result of the LLM was double-checked and all sources were manually verified. We used a custom annotation viewer to easily compare annotations side-by-side, as shown in \Cref{fig:annotation_viewer}.
\end{enumerate}

\begin{lstlisting}[caption={Web Research Prompt: Sent to the model with web search enabled to gather primary sources about the agent.}]
Research the specific AI agent product "{agent_name}" to verify information about: {column_name}

IMPORTANT: Focus ONLY on this specific agent product, NOT:
- The underlying foundation models (e.g., GPT-4, Claude) - unless asking about model specifications
- The parent company's general information or other products
- Generic AI industry information
- Related but different products from the same company

Field guidance: {field_description}

Current annotation to verify:
{current_annotation}

Known resources for this agent: {resources}

Search for PRIMARY SOURCES specifically about "{agent_name}":
- Official product documentation and help pages for this agent
- Product-specific press releases and announcements
- Developer blogs about this specific agent
- Trust center pages specific to this product
- Product landing pages and feature descriptions

Provide a concise research summary with URLs to sources found. Only include information directly about "{agent_name}".
\end{lstlisting}

\begin{lstlisting}[caption={Structured Verification Prompt: Formats research findings into structured JSON output.}]
Based on the research below, provide a verification of the annotation for the specific agent product "{agent_name}" - field "{column_name}".

## Field Guidance
{field_description}

## Current Annotation
{current_annotation}

## Research Findings
{research_context}

## Critical Instructions
- Focus ONLY on this specific agent product "{agent_name}"
- Do NOT include information about underlying foundation models unless specifically relevant to the field
- Do NOT include general company information unless specifically relevant
- Provide object-level findings with inline citations
- Use "None found" if no info exists, "Not applicable" if field doesn't apply, "UNSURE" if uncertain

Provide your verification result.
\end{lstlisting}

\begin{lstlisting}[caption={System Message: Accompanies the structured verification prompt to enforce output format.}]
You are a research assistant verifying annotations for the specific AI agent product "{agent_name}".
Focus ONLY on this agent product - not underlying models, parent company general info, or related products.

OUTPUT FORMAT: Provide a concise bullet-point list of findings, NOT paragraphs.
Each bullet should be a single factual finding with an inline source link.

Example format:
- Feature X is supported [link](https://example.com/docs)
- Pricing starts at $20/month [link](https://example.com/pricing)
- No information found on Y

Return a JSON object with these exact keys:
- verified_annotation: string (concise bullet-point list of findings, each with inline markdown link)
- confidence: string ("High", "Medium", or "Low")
- confidence_explanation: string (why you rated confidence this way)
- notes: string (additional notes for research team)
\end{lstlisting}

\subsection{Correction Request to Companies}
To ensure accuracy, we contacted each company with agents in the Index on December 12, 2025, providing a view-only link to our draft annotations and inviting corrections by January 11, 2026. We committed to maintaining confidentiality of all correspondence while noting that the final database would be released publicly. Below is the email template used:

\begin{quote}
We are a team of academic researchers from MIT, Harvard, Stanford, and other universities, updating the AI Agent Index published by Casper et al. (2025), which documented 67 AI agents and has become a widely-cited resource in the field.

The 2025 edition focuses on a deeper analysis of publicly available information on approximately 30 particularly significant agentic systems, including [AGENT NAME(S)]. 
To ensure the accuracy and completeness of our work, we kindly request your feedback on the information we collected regarding [AGENT NAME(S)] (view-only link to the draft sheet). Note that “None found” means we are unable to find any public information on this field. You can share your feedback by replying to this email at [redacted] by January 11th, 2026.
All correspondence in this email thread will remain strictly confidential and will be used solely to improve the accuracy of our database. The finalized database (including any changes we make based on your feedback) will be released publicly along with the research paper. Please refrain from sharing the database prior to its publication.

Your expertise is key to ensuring the rigor and reliability of our work, and we deeply value your input. If you have any questions, feel free to reach out. 

\end{quote}

\section{Public Interest Methodology}\label{app:search}
For search volume metrics, we consider combinations of product and company name as well as simpler terms such as just product name (if unambiguous) or company + \verb|AI agent|, etc.

For GitHub, we consider the highest number of stars in 2025. If an agent product has a GitHub repository, this does not imply that it is open-source. The open-source annotation field was validated separately.

\subsection{LLM Prompts for Generating Search Terms}
We used OpenAI GPT-5.2 with web search to create a list of possible search terms for each \verb|company| and \verb|product| combination in two steps. 

First, we used OpenAI GPT-5.2 with web search to research the product. Below is the prompt used for this.
\begin{lstlisting}    
Research {company} {product} and provide a concise summary covering:
1. What it is (e.g., AI coding assistant, browser, automation platform)
2. Primary use case and key features
3. What makes it unique or distinguishing characteristics

Search the web for the most current and accurate information.
Keep the response focused and factual, under 150 words.
\end{lstlisting}

Second, we used OpenAI GPT-5.2 to generate a list of relevant search terms based on the product background from step 1. Below is the prompt used for this.
\begin{lstlisting}    
Given this information about an AI agent product:
- Company: {company}
- Product Name: {product}
- Context: {context}
{existing_terms_str}

Generate NEW unambiguous search terms that meet these STRICT requirements:

IMPORTANT: Only generate NEW terms that are NOT in the existing list above. Do not duplicate any existing terms.

1. MUST INCLUDE BASIC COMBINATIONS (if not already existing):
   - "{company.lower()} {product.lower()}" (if not in existing)
   - If company and product names are DIFFERENT, also include "{product.lower()} {company.lower()}" 
   (reversed order, if not in existing)
   - If company and product names are the SAME, skip the reversed term (no duplicates)

2. STANDALONE PRODUCT NAME (when applicable):
   - IF the product name is sufficiently unambiguous on its own, include it
   - GOOD: "chatgpt agent" (specific, won't confuse with general ChatGPT)
   - GOOD: "windsurf" (unique enough to be unambiguous)
   - BAD: "comet" (too generic, could mean space comet)
   - BAD: "atlas" (too generic, could mean geography atlas)
   - BAD: "north" (too generic, could mean direction)
   - Only include if you're confident it won't be confused with other meanings

3. CONCISE: Use simple keyword combinations, NOT descriptive phrases
   GOOD: "perplexity comet", "comet browser", "perplexity comet browser"
   BAD: "AI-powered Comet chatbot for accurate answers"

4. SPECIFIC: Terms must clearly identify THIS SPECIFIC PRODUCT, not the company generally
   GOOD: "chatgpt agent mode", "openai chatgpt agent"
   BAD: "ai agent" alone, "openai" alone (too general, refers to company or generic product)

5. UNAMBIGUOUS: Cannot be confused with other meanings
   GOOD: "perplexity comet browser" (not space comet)
   BAD: "comet" alone (could mean astronomical comet)

6. INCLUDE COMPANY + PRODUCT CATEGORY: Include combinations of company name + product type/category
   GOOD: "perplexity browser" (for Perplexity Comet which is a browser)
   GOOD: "openai coding agent" (for ChatGPT Agent which is a coding agent)
   GOOD: "google ai agent" (for Project Mariner which is an AI agent)
   - These are important because users often search for "company + what it is"

7. NATURAL: Phrases a user would actually search on Google
   - Combine company name + product name
   - Combine company name + product category (e.g., "perplexity browser")
   - Avoid duplicating words when combining company name + product name 
   (e.g., "openmanus" is better than "openmanus openmanus")
   - Combine product name + key identifier (e.g., "browser", "agent", "assistant")
   - Vary word order for natural search patterns

It is ESSENTIAL that the generated search terms are unambiguous and specific to the product. 
It needs to just be about the product itself, not about related subquestions. 
ONLY include search terms that are directly to the product.

Return ONLY a JSON array of strings with NEW terms only (not in existing list), no explanations.
If all good terms already exist, return an empty array [].
Example: ["new term 1", "new term 2", ...]
\end{lstlisting}

\subsection{Public Interest Analysis}

\begin{figure}[H]
    \centering
    \includegraphics[height=0.4\textheight,width=\linewidth]{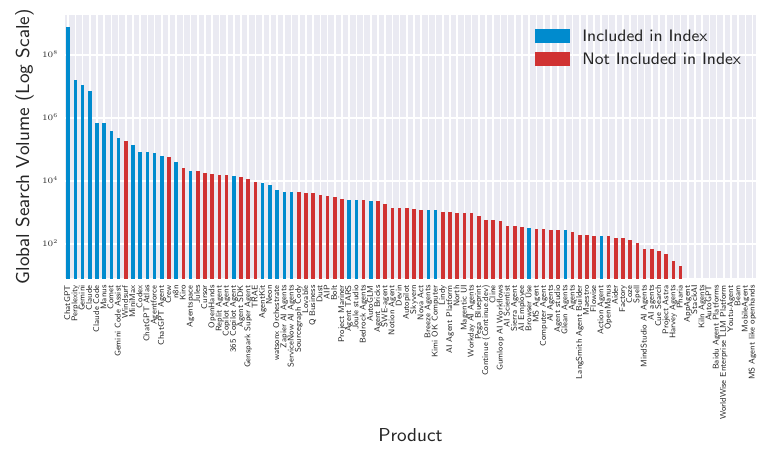}
    \Description{Ranked bar chart of global search volume on a logarithmic scale. Blue bars are included products and red bars are not included. The left side contains a few much taller bars, followed by a long descending tail of shorter bars across many products.}
    \caption{Global search volume (log scale) for each AI agent product. Blue indicates inclusion in the Index.}
    \label{fig:search_volume_statistics}
\end{figure}

\begin{figure}[H]
    \centering
    \includegraphics[width=\linewidth]{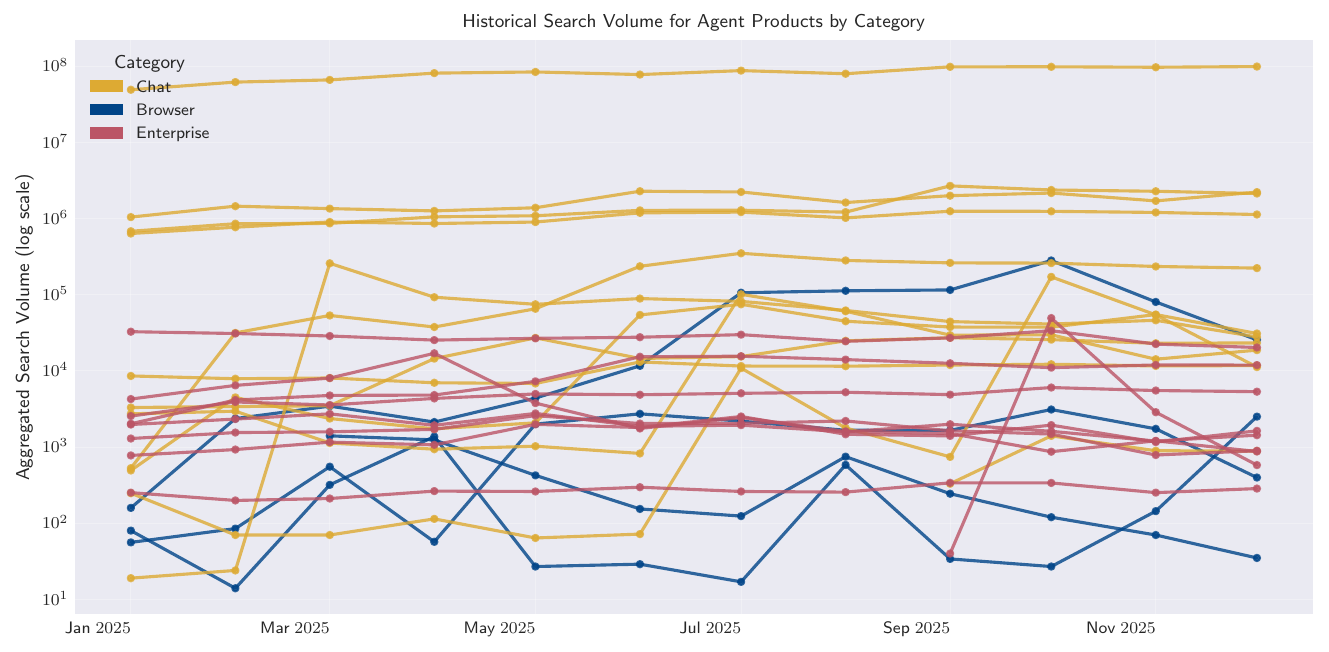}
    \Description{Line chart of monthly search volume for indexed products in 2025. The horizontal axis runs from January to December 2025 and the vertical axis is logarithmic. Chat products, shown in yellow, occupy most of the highest lines, while browser and enterprise products are generally lower with a few temporary spikes.}
    \caption{Monthly search volume (log scale) for the most popular term for each AI agent product included in the Index over 2025, colored by agent category (\ulchat{chat}, \ulent{enterprise}, and  \ulbrowser{browser}).}
    \label{fig:search_volume_historical}
\end{figure}

\begin{figure}[H]
    \centering
    \includegraphics[width=1\linewidth]{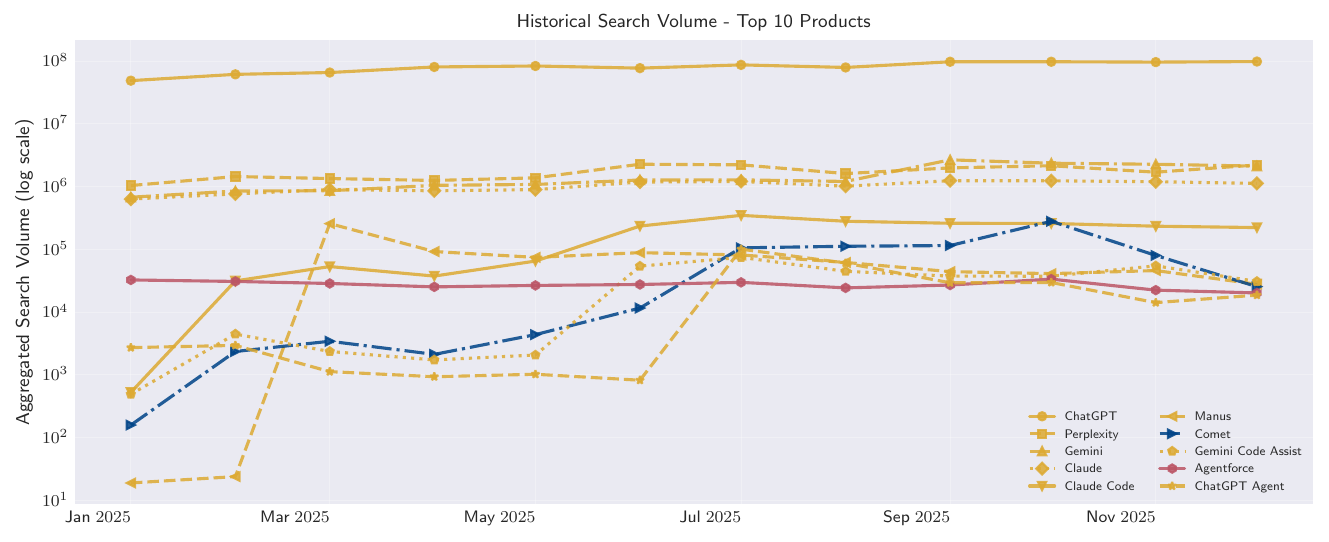}
    \Description{Line chart of monthly search volume for the top ten products in 2025. One chat product, ChatGPT, stays highest across the year near the top of the logarithmic scale. Several other chat products form a middle band, while Manus, Comet, Agentforce, and ChatGPT Agent rise sharply at selected points before dropping back.}
    \caption{Monthly search volume (log scale) for the 10 most popular AI agent products, colored by agent category (\ulchat{chat}, \ulent{enterprise}, and  \ulbrowser{browser})}
    \label{fig:search_volume_historical_top_products}
\end{figure}

\begin{figure}[H]
    \centering
    \includegraphics[width=1\linewidth]{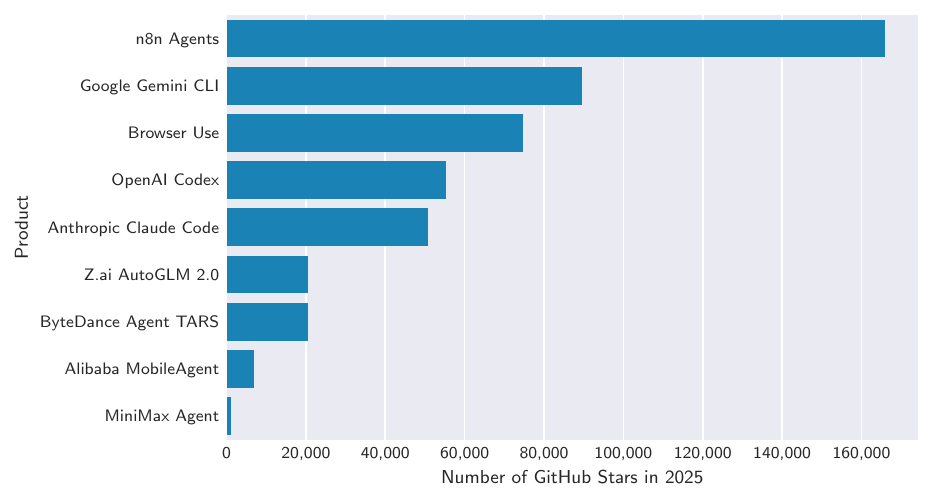}
    \Description{Horizontal bar chart of GitHub stars in 2025 for agent-related repositories. n8n Agents has the longest bar by a wide margin, followed by Google Gemini CLI and Browser Use. The remaining repositories decline in steps to much shorter bars for Alibaba MobileAgent and MiniMax Agent.}
    \caption{GitHub stars for repositories associated with the agent products.}
    \label{fig:github_stars_analysis}
\end{figure}

\section{Supplementary Materials for Case Studies}\label{app:case_studies}

\begin{figure}[H]
    \centering
    \includegraphics[width=1\linewidth]{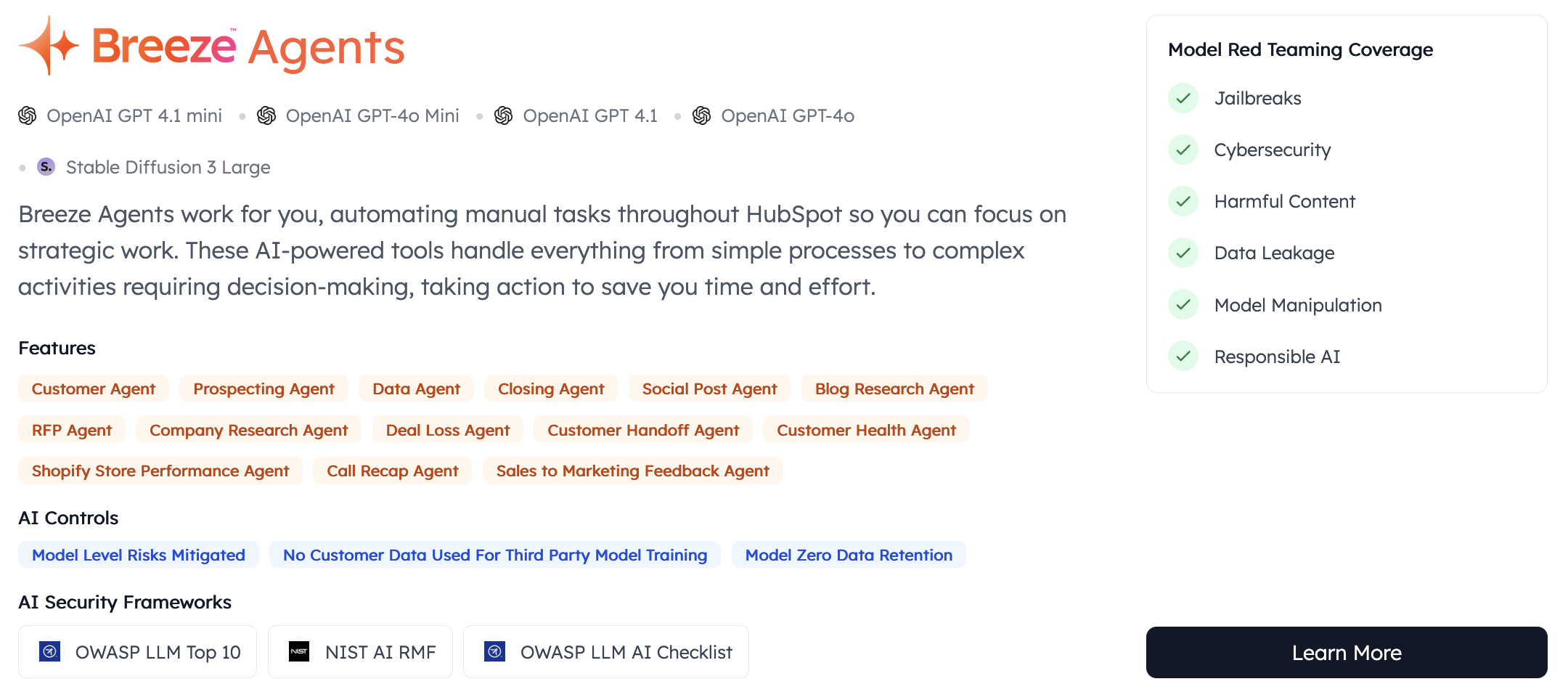}
    \Description{Screenshot of the Breeze Agents page from HubSpot. The page has a large Breeze Agents heading, a list of supported model names, feature chips for different agent types, and AI control labels such as no customer data used for third-party model training. A right sidebar titled model red teaming coverage lists six checked items: jailbreaks, cybersecurity, harmful content, data leakage, model manipulation, and responsible AI.}
    \caption{Details on the safety features and red teaming methodology for AI agents are limited. Screenshot of the ``Model Card'' for HubSpot's Breeze Agents taken from \url{https://trust.hubspot.com/ai}.}
    \label{fig:breeze_safety_screenshot}
\end{figure}
\end{document}